\documentclass[12pt]{article}
\usepackage{fullpage,amsmath,amssymb,amsfonts,natbib,xcolor,url,graphicx,subcaption,hyperref,soul}
\setstcolor{red}
\usepackage{float}
\def\lo{\mathcal L}
\def\hi{\mathcal H}
\def\cov{\text{cov}}
\def\var{\text{var}}

\begin{document}
\title{\sc
%\normalsize Heat Waves and Warming Trends: \\ 
The Economic Impact of  Low- and High-Frequency  Temperature Changes}
\author{Nikolay Gospodinov\footnote{Federal Reserve Bank of Atlanta, Email: nikolay.gospodinov@atl.frb.org} \and Ignacio Lopez Gaffney\footnote{Department of Economics, Yale University and Yale Law School, Email: ignacio.gaffney@gmail.com} \and Serena Ng\footnote{Corresponding author: Department of Economics, Columbia University and NBER. 420 W. 118 St. New York, NY 10027. Email: serena.ng@columbia.edu \smallskip \newline We would like to thank the Editor, two anonymous referees, the participants at the 2025 Conference on The Macroeconomic and Financial Aspects of Climate Change (Mexico City), and Luis Sarmiento (discussant) for very useful comments and suggestions. This work is supported by  the National Science Foundation (SES: 2018369). The views expressed in this paper are those of the authors and do not necessarily represent those of the Federal Reserve Bank of Atlanta or the Federal Reserve System.}}

\date{\today}
\maketitle
\thispagestyle{empty}
\setcounter{page}{0}
\bibliographystyle{harvard}
\begin{abstract}
\noindent 
%Temperature data have low- and high-frequency variations that may have distinct impacts on economic outcomes. Analyzing data from a panel of 48 states in the U.S., and a panel of 50 countries, we find  slowly evolving, low-frequency components with periodicity greater than 15 years. These components have a common factor that trended up around the same time that economic growth slowed. Panel  regressions using U.S. data  fail to find a statistically significant impact of  low-frequency temperature changes on growth, though the impact of high-frequency temperature changes is marginally significant. However, using  the international panel (which includes several European countries), we find that a 1°C increase in the low-frequency component is estimated to reduce economic growth by about one percent in the long run. Though the first-order effect of high frequency changes is not statistically significant in this data, a smaller non-linear effect is detected. Our estimation and inference procedures control for common, business cycle variations in output growth that are not adequately controlled for by an additive fixed effect specification. These findings are corroborated by time series estimation using data at the unit and national levels.

\noindent Variations in the low- and high-frequency components of temperature  may have distinct impacts on economic outcomes. Parametric and non-parametric  estimates from three panels of data all find significant heterogeneity in the relative importance of the two components, but there is clear evidence in each panel of a common,   slowly evolving low-frequency factor  that is highly correlated with the low-frequency factor of economic activity.  In regressions that quantify the output effects of the components, we find that  one-way clustered standard errors often lead to size distortions, and that an additive fixed effect specification does not adequately  control for common time effects. Using bootstrap inference to assess estimates from our preferred interactive fixed effect specification, we only find  a marginally significant  effect of the high-frequency component on growth in the U.S. panel. However,  the effect of the low-frequency component is significant in  the European and International panels,  suggesting that the increase in the low-frequency temperature component over the post-1980 period is associated with a reduction in economic growth of approximately 1.3 percentage points.
%following  a 1°C increase in the low-frequency component.    
The findings are corroborated by time series estimation using data at the unit and national levels. 

\bigskip

\end{abstract}
\newpage
\baselineskip=18.0pt
\section{Introduction}

 Average global temperatures have increased by 0.06°C per decade since 1850 – or about 1.11°C in total – with warming occurring three times faster since 1982 (0.20°C per decade).\footnote{ \url{ https://www.climate.gov/news-features/understanding-climate/climate-change-global-temperature} }  Non-seasonal increases in temperature can arise due to heat waves or due to El Ni\~{n}o and La Ni\~{n}a, which shift climate patterns in the Pacific Ocean roughly every 3-7 years. Variations can also occur due to solar cycles that flip the sun's magnetic field every 11 years,  as well as alternating periods of cool-damp and warm-dry weather related to 35-year Bruckner cycles.\footnote{ \url{https://en.wikipedia.org/wiki/Eduard_Bruckner}.} The possible existence of temperature variations with different  periodicities  will affect how we model the economic impact of environmental changes and how we design mitigation policies (\citet{carleton-duflo:24}).  It is also of economic interest to understand how  agents distinguish the different types of  variations from the data.

 The  literature often treats  weather – atmospheric conditions over short periods of time – as conceptually  distinct from climate, represented by moments  of  weather variables computed over a long sample (e.g., \citet{ipcc-2014}).  However, there is another distinction of interest: temperature variations that are slowly evolving with effects that are persistent,  versus those that are more volatile and whose effects are short-lived.  We focus on one weather variable – temperature – and model observed  temperature  $X$  as the sum of two unobserved components: one of `low' frequency, denoted by $\lo_X^0$ (akin to climate), generating persistent effects, and one of `high' frequency, denoted by $\hi_X^0$ (akin to weather), generating short-lived effects. In a stationary environment, $\lo_X^0$ would be the  constant, long-run mean, and $\hi^0_X$ would be a series of deviations from this constant. To capture non-stationarity,  we  allow $\lo^0_X$  to be a slowly evolving process, possibly with long memory.

In this paper, we show that low- and high-frequency variations may transmit differently to economic activity. Our analysis has two distinct features. First, we characterize the dynamics and factor structure of the low-frequency variations in temperature. This informs our approach to investigate if low- and high-frequency variations in temperature have a differential impact on economic outcomes. Second, our regression analysis pays attention to the treatment of unobserved heterogeneity and common time effects. To this end, we use an interactive fixed effects model, combined with bootstrap inference that accounts for the estimation uncertainty of the unobserved factors and the clustered data structure. We consider 3 panels – U.S. (48 states), European (20 countries) and international (50 countries) panels – which allow us to assess the effects across different geographical regions.

As  $\lo^0_X$ is not observed, we first explore parametric and non-parametric ways of estimating it.   
% We begin by  fitting a parametric  unobserved components model with long memory  to $X_i$ for each $i=1,\ldots, N$ in the panel and find significant spatial differences in the variability of $\hat \lo_{Xi}$ relative to $\hat \hi_{Xi}$. 
Regardless of the estimator, we find in all three panels that  the low-frequency component   has  been trending up  in recent decades but displays heterogeneous features across geographic units.    
%  These features are corroborated by  three other trend estimators used in economic analysis, and by data at the national level.  A  Monte Carlo exercise   calibrated to temperature data  finds that 
% with the `right'  hyperparameter, all methods provide a similar approximation to  $\lo^0_X$.  
%In the regression analyses, we use the procedure of \citet{mueller-watson:18}  to construct  $\hat\lo_X$ and $\hat\hi_X$,  which are orthogonal by construction. 
Regressions using the U.S. panel over the 1964--2023 sample find a significant and negative  effect of the low-frequency component, but only under the assumption that there are no common year effects. Even in this model, the estimates deemed statistically significant according to one-way clustered standard errors, often used in this literature, become insignificant using two-way clustering. Models that are better at controlling for common year effects are also not significant. Therefore, even though $\hat \lo_X$ increased by 1.72°C  between 1964 and 2023, it does not seem to have had a meaningful impact on economic growth in the United States. The evidence from a panel of 20 European countries, and an international panel of 50 countries with wide geographic coverage and variation in incomes, is more supportive of a  negative effect of $\hat\lo_X$ on GDP growth, $\Delta Y$. 
% Based on dynamic specifications of interactive fixed-effect regressions, we estimate a coefficient for $\hat \lo_X$ of around $-0.75$ in the European and international panels. Considering that the population-weighted average of  $\hat \lo_X$  in Europe has increased by 1.48°C in the 38 years between 1980 and 2018, these estimates translate into an annual impact on GDP growth of about 0.03 percentage points. 
Based on a static interactive fixed-effect regression, we estimate a coefficient for $\hat \lo_X$ of around $-0.9$ in the European panel. Considering that the population-weighted average of  $\hat \lo_X$  in Europe has increased by 1.48°C between 1980 and 2018, these estimates translate into an annual reduction of GDP growth of about 0.035 percentage points.

As for the high-frequency component, the first-order impact  is negative and marginally significant in the U.S. panel, but not in the European or international panel. However, there is suggestive evidence of a nonlinear effect in  both panels when the marginal effect of $\hat \hi_X$ is allowed to vary with $\hat \lo_X$. The data thus suggest that the output effect of $\hat \lo_X$ is negligible in the U.S. while that of $\hat\hi_X$ is negligible in the Europe. Put differently, U.S. output growth is associated with changes in the high-frequency component with effects that are short-lived, while growth at the international level is associated with changes in the low-frequency component with effects that are long-lasting. Further analysis is needed to understand the economic  and environmental factors that underlie this difference. 

In addition to  panel regressions, we also perform time series regressions using  unit-level and aggregate data. These   regressions lead to the same conclusion that the long-run impact of the low-frequency component in the U.S. on growth is not statistically different from zero. The aggregate estimate for  European countries also corroborate the panel data results of a significant negative relationship between the low-frequency components of temperature and growth, although the estimated effect from the aggregate time-series regression is even larger than the one obtained from the panel regressions.  As there is no need to control for heterogeneity in this setting, it may be appealing to look for aggregate evidence from time series regressions, especially as we gain more time series observations going forward.

We view our findings as complementary, and broadly consistent, with the growing literature that examines the effect of temperature (or climate, more generally) on economic growth or potential output (see, for example, \cite{dell-jones-olken:12}, \cite{burke-emerick:16}, \cite{donadelli-et-al:17}, \cite{colacito-hoffman-phan:19}, \cite{leduc-wilson:23}, \cite{bilal-rossi-hansberg:23}, \cite{bilal-kanzig:24}, \cite{natoli:26}, \cite{nguyen:26}, among others). Direct comparisons with the results in other studies is difficult not only because of the different methodologies and data samples, but because our static and dynamic panel regressions focus on  how the high- and low-frequency components of temperature affect economic activity across geographic regions while controlling for common variations. 
 
The rest of the paper proceeds as follows. Section \ref{sec:Sect2}  uses a dynamic model to  motivate the decomposition of $X$  into high- and low-frequency components. It also reviews several non-parametric estimators of these latent processes, and characterizes the common and idiosyncratic  variations in $X$ and $\Delta Y$ for U.S. data. Section \ref{sec:Sect4} uses regressions to estimate the average and state/country level effects of the components of $X$ on $\Delta Y$. Section \ref{sec:Sect5} concludes. 

\paragraph{Data:}  In this paper, we use data from three panels. The U.S. panel consists of data for the 48 contiguous states. Temperature data is available for the 1895-2023 sample, while the growth rate of real per capita Gross State Product (GSP) is available for the 1964-2023 sample.
%\footnote{We also considered quarterly (seasonally adjusted) data for the 1948Q1--2023Q4 period. The findings are qualitatively similar.} 
To facilitate comparisons with the international data, the temperature series for the U.S. are converted from °F (Fahrenheit) to °C (Celsius). The international panel consists of data for 50 (relatively developed) countries. Temperature data are available for the 1901-2022 sample, while the growth rate of real per capita GDP is available for the 1953-2018 sample. A European panel (20 countries) is obtained as a subset of this international panel. We construct temperature series as deviations from their pre-1980 mean. This data transformation  allows all series to have approximately the same level, and facilitates comparison across geographic units. We will often refer to `national' or `aggregate' temperature series, which are defined as population-weighted averages of unit temperature series. A detailed description of the data is provided in \ref{app:AppendixC}. 
%Table \ref{tbl:summary} in \ref{app:AppendixC} summarizes the data properties of the three panel datasets used in the paper.

\section{Modeling  $\lo^0$} 
\label{sec:Sect2}
The notion of low-frequency variations is important in our analysis, so it is useful to be clear on what we are studying. A series with  $T$ observations can emerge from different configurations of the sampling frequency  and span of  the data. For a given  sampling frequency and span,  our point of departure is that a time series  can be expressed as a sum of  orthogonal components, each isolating variations with a different periodicity (e.g., \citet[Chapter 6]{harvey-white}).  Component $j$ has periodicity  $p_j=\frac{2\pi}{\omega_j}$ if it repeats every $p_j$ periods, where $\omega_j\in[0,\pi]$ is the  frequency at which the spectral density has notable mass.  For economic time series, such as the inflation  rate, focus is usually on the business cycle component, which has periodicity between 6 and 32 quarters. 

Temperature data, likewise, have variations at different frequencies that are of interest.  We work with annual data, which are void of variations at seasonal frequencies, to focus on the low-frequency trend $\lo^0_X$ (i.e., the trend attributable to variations at zero or near-zero frequencies). With some abuse of terminology, we will refer to $X-\lo^0_X\equiv \hi^0_X$  as the high-frequency component, even though these variations may still be persistent. The crucial distinction is when $\omega_j$ is near zero, the periodicity is high. 

Many regressors considered in the literature can be seen as proxies for $\lo_X^0$ and $ \hi_X^0$.   For example,  $\Delta X$, the number of heating-degree days and deviations from a 30-year rolling  mean  can be viewed as estimates of $ \hi^0_X$, while a rolling mean is an estimate of $\lo^0_X$.\footnote{See also \citet{wu-etal:07}, \citet{mills:07}, \citet{mudelsee:review}, \citet{kaufmann-etal:13} for an analysis of non-stationarity.} %\citet{bastien-olvera:22} employ a low-pass filter with different cutoffs to approximate $\lo^0_X$, which is similar in spirit to our analysis.
%As discussed in \citet{mueller-watson-ecma:08}, their cosine transforms produce nearly ideal low-frequency realizations.
\citet{zhang:07} use a 40-year Butterworth filter to look for  patterns between  wars and temperature changes, while \citet{tol-wagner:10} use a 10-year Hamming window. \citet{bastien-olvera:22} remove variations with periodicity up to 15 years. \citet{hsiang:16} uses the band-pass filter with different cut-offs to  focus on higher frequency  variations in corn yields. However, the effectiveness of these  procedures  has not been studied in a systematic manner. Trend–cycle decomposition is, however, a well-studied problem in the economics literature. We leverage this knowledge to analyze the low-frequency component of temperature data by first using a parametric model to gain insights about its underlying persistence, and then resorting to other non-parametric ways to isolate it. As will be seen below, our estimate of $\lo^0_X$, based on a sample of $T=129$ years, has a periodicity of approximately 32 years.

\subsection{An Unobserved Components Model with Long Memory}

For a generic variable $Z_t$ observed for $t=1,\ldots, T$,  an unobserved components (UC) model  assumes 
\[Z= \lo^0+\hi^0.\] 
We observe $Z$ but not the two latent components.  The standard UC model parameterizes  $\hi^0$ as a covariance stationary process and $\lo^0$ as an $I(d)$ process with  $d$ taking on integer values of  one or two. When $Z$ is economic output,  \citet{beveridge-nelson-jme} and \citet{kuttner:94} associate $ \lo^0$ with  potential output,  while \citet{clark:87} associates $\hi^0$  with the business cycle. \citet{arino-marmol} and, more recently,  \citet{hartl-uc} relax the assumption that $d$ is integer valued. This more flexible model entertains the possibility that temperature data has long memory, as documented in  \citet{mills:07}, \citet{gil-alana:22}, \citet{yuan-fu-liu:14}, among others.

The low- and high-frequency components are assumed to  evolve according to
 \begin{eqnarray*}
  (1-L)^d \lo^0_{t}&=&\epsilon_{\lo,t}, \quad\quad\quad\quad\quad  \epsilon_{\lo,t}\sim (0,\sigma^2_\lo),\\  
    \hi^0_{t}&=&a(L) \epsilon_{\hi,t}, \quad\quad\quad  \epsilon_{\hi,t}\sim (0,\sigma^2_\hi), 
 \end{eqnarray*}
  where $a(L)=1+a_1L+\ldots a_pL^p$ is a lag polynomial of order $p$. Both $\lo^0$ and $\hi^0$ depend on the parameters    $(\sigma_\hi,\sigma_\lo, d, a_1, \ldots, a_p)$.  \citet{nguyen:26} also estimates the low-frequency component for temperature with the restriction that $d=1$.
  
%   Let $b(L)=a(L)^{-1}=\sum_{j=0}^\infty b_j L^j$. Assuming that  $\sum_{j=0}^\infty |b_j|$ is bounded, and bounded away from zero,  \citet{hartl-uc} shows that optimal filtering yields 
%   \begin{eqnarray*}
% \lo_{t:1}(Z_{t:1},\psi)&=&  (B_t'B_t + \nu S_t'S_t)^{-1} B_t'B_t Z_{t:1},\\
%  \hi_{t:1}(Z_{t:1},\psi)&=& \nu (B_t'B_t + \nu S_t'S_t)^{-1} S_t'S_t Z_{t:1},
% \end{eqnarray*}
% where   $\nu=\sigma^2_{\hi}/\sigma^2_\lo$,
%  $S_t$ and $B_t$ are  upper triangular Toeplitz matrices.\footnote{The $S_t$ matrix is a function of $(\pi_0(d), \pi_1(d),\ldots, \pi_{t-j}(d))$, $\pi_j(d)=\frac{j-d-1}{j} \pi_{j-1}(d)$ starting with $\pi_0(d)=1$,  and the $ B_t$ matrix has $(b_0, b_1, \ldots,b_{t-j})$ in the $j$th-row, respectively.}  Both $\lo^0$ and $\hi^0$ depend on the known  parameters    $(\sigma_\hi,\sigma_\lo, d, a_1, \ldots, a_p)$. 
 
 Let $\hat \lo_X$ and $\hat \hi_X$ be estimates of the components of temperature $X$  where the parameters are estimated by maximum likelihood (for details, see \citet{hartl-uc}).
 % \footnote{\citet{hartl-uc} allows $a_j$ to depend on parameters $\varphi$.  For our purpose, it suffices to remove this generality.}  
 Using U.S. data from 1895 to 2023, which is the longest sample available, our  estimation with $p=1$  finds $\hat\sigma_\lo$ to be  small. This is concerning because even in the standard UC model when $d$ is fixed at $1$, identification is challenging  when $\sigma_\lo/\sigma_\hi$ is small.   Therefore, we fix $\sigma_\lo$ to take on values between 0.01 and 0.3 and estimate only the remaining parameters, $(\sigma_\hi,d,a_1)$. We find that while the  likelihood is maximized  at $\sigma_\lo=0.02$, it is  similar for any $\sigma_\lo\in[0.02,0.2]$.  Though $\hat \lo_X$  is more variable when $\sigma_\lo=0.2$ than when $\sigma_\lo=0.02$, it has a similar autocorrelation. Of more interest is the significant dispersion  in $\hat\sigma_\hi$ across states and that $\hat d$ always exceeds 0.5. This makes $\hat \lo_X$ non-stationary  for any $\sigma_\lo\in [0.02, 0.2]$ suggesting that we can analyze any series with $\sigma_\lo$ in this range and extract similar conclusions. Hereafter, we will denote a UC model with $\sigma_\lo$ fixed at $\omega$ as UC$\omega$. It is also convenient to define the ratio $\nu=\sigma^2_{\hi}/\sigma^2_\lo$. 
 
\begin{table}[ht]
%% plotd.tex
\caption{UC Estimates of 48 U.S. States Ordered by $\sqrt{\hat \nu}=\hat\sigma_\hi/ \sigma_\lo$}  
\label{tbl:sbs_plot_d}
\vspace*{-.2in}
\begin{center}
% 1895-2023 \qquad $T=129$ \qquad $\sigma_\lo=0.2$ 
\begin{tabular}{lllll|llll|llllll}
& State & $\hat d$ & $\Delta^T \hat \lo_X $ & $\hat \sigma_\hi$  
& State & $\hat d$ & $\Delta^T \hat \lo_X $ & $\hat \sigma_\hi$ 
& State & $\hat d$ & $\Delta^T \hat \lo_X $ & $\hat \sigma_\hi$  \\ \hline
& \multicolumn{4}{c}{Bottom third} & \multicolumn{4}{c}{ Middle Third} & \multicolumn{4}{c}{Top Third} \\ \hline

&        \bf{FL} &1.036 &1.745& 0.401&        \bf{WA} &0.869 &0.890& 0.552&        UT &1.000 &1.420& 0.641\\
 &        \bf{CA} &1.023 &1.557& 0.413&        TN &0.950 &0.821& 0.556&        {VT} &1.061 &1.791& 0.647\\
 &        AZ &0.997 &1.501& 0.435&        \bf{MA} &1.092 &2.050& 0.557&        WY &0.957 &1.176& 0.662\\
 &        NM &1.014 &1.421& 0.446&        \bf{CT} &1.060 &1.928& 0.560&        OH &0.981 &1.233& 0.669\\
 &        \bf{SC} &0.971 &1.003& 0.473&        \bf{MD} &1.047 &1.711& 0.563&        MO &0.858 &0.742& 0.701\\
 &        \bf{LA} &0.940 &0.917& 0.480&        WV &0.978 &0.983& 0.567&        IN &0.910 &0.887& 0.710\\
 &        \bf{NC} &1.001 &1.108& 0.485&        \bf{NJ} &1.077 &2.100& 0.567&        KS &0.868 &0.754& 0.716\\
 &        NV &0.986 &1.242& 0.487&        CO &0.933 &1.110& 0.569&        MI &1.016 &1.544& 0.726\\
 &        \bf{GA} &1.008 &1.048& 0.488&        PA &1.015 &1.555& 0.575&        MT &0.935 &1.060& 0.746\\
 &        \bf{AL} &0.984 &0.817& 0.505&        \bf{NH} &1.062 &1.862& 0.582&        IL &0.911 &0.972& 0.761\\
 &        \bf{MS} &0.962 &0.812& 0.507&        \bf{NY} &1.051 &1.870& 0.588&        NE &0.904 &0.856& 0.789\\
 &        \bf{TX} &0.965 &1.155& 0.509&        AR &0.838 &0.517& 0.596&        WI &0.948 &1.132& 0.820\\
 &        \bf{OR} &0.909 &1.061& 0.512&        \bf{ME} &1.054 &1.805& 0.598&        IA &0.806 &0.576& 0.821\\
 &        \bf{VA} &1.020 &1.397& 0.515&        OK &0.867 &0.520& 0.619&        SD &0.873 &0.818& 0.882\\
 &        \bf{RI} &1.098 &2.080& 0.544&        KY &0.977 &0.962& 0.620&        MN &0.952 &1.166& 0.916\\
 &        \bf{DE} &1.047 &1.791& 0.546&        ID &0.882 &0.871& 0.633&        ND &0.926 &1.033& 0.983\\
 \hline

\hline
\end{tabular}
\end{center}
{\footnotesize The sample period is 1985--2023 ($T=129$). $X$ is the deviation of temperature (in Celsius) from the pre-1980 mean. $\Delta^T \hat \lo_X=\hat \lo_{X,2023}-\hat \lo_{X,1895}$, $\hat \lo_X$ is estimated by the UC model with $\sigma_\lo=0.1$, $\hat d$ is the long memory parameter, and $\hat \sigma_\hi$ is the innovation variance of $\hi_X$ parameterized as an AR(1). States with ocean coastline are in bold font.}

\end{table}

 Table \ref{tbl:sbs_plot_d} shows results   ordered by $\hat \sigma_\hi$ for $\sigma_\lo=0.1$. Also reported is $\Delta^T \lo_X=\hat \lo_{X,T}-\hat \lo_{X,1}$, the change in the low-frequency component over the sample. To facilitate visualization of spatial differences, states with ocean coastline are presented in bold font. The states with the largest $ \hat\sigma_\hi$ (e.g., ND, MN, SD, IA, WI) are located in the middle of the country. The states with the lowest $\hat\sigma_\hi$ (e.g., FL, CA, SC, LA, NC) are warm and coastal. The population-weighted average of the parameters $(\hat d,\hat a_1,\hat \sigma_\hi,\hat \nu)$  are (1.039, 0.137, 0.974, 24.990), with cross-sectional standard deviations of (0.056, 0.107, 0.263, 16.271). The national estimates are (1.019, 0.201, 0.628, 9.886). Both the state-level and national estimates find weak persistence in $\hat \hi_X$.  The national  $\hat\sigma_\hi$ of $0.628$ is lower than the aggregate estimate of $0.974$, as averaging the data prior to estimation reduces noise. Note that with $\sigma_\lo=0.1$, the $\hat d$ estimates  are all below $1.5$ at the state level, while $\hat d$ is close to $1$ at the national level. This aligns the national data closely with the standard UC model, where $d=1$. 

 The distinctive feature  in Table \ref{tbl:sbs_plot_d} is that of $\Delta^T \hat\lo_X$ is  much larger in the coastal states.   \citet{gadea-rivas-gonzalo:25} used the distribution of  temperature to quantify changing  warming  patterns and find that 84\% of the  states in the U.S. show significant warming, much higher than data on the mean of temperature suggests.   \citet{bilal-rossi-hansberg:23} find, using  county level data,  that the frequency of severe storms has risen particularly fast in the coastal regions. We  focus on the low-frequency variations and also arrive at the  conclusion of heterogeneity in the nature of temperature changes across states.

An interesting aspect of the UC model is that  when $t=T$, $d=2$, and $a(L)=1$, $\hat \lo_X$ is equivalent to the  Hodrick-Prescott (HP) filter, with $\nu$ serving as a smoothing parameter denoted as $\lambda$.  For quarterly economic data, a $\lambda$ of $1600$ is found to  roughly isolate  business cycle variations (i.e., those with  periodicity  between 6 and 32 quarters). \citet{uhlig-ravn} suggest setting $\lambda$ to 6.25 when using annual data to obtain a cyclical component with similar spectral properties. Our  $\hat d$ estimates are well below $2$, and the  $\hat\nu$ estimates are well above  $6.25$, suggesting that a $\lambda$ suitable for economic data is not appropriate for analyzing the low-frequency component of $X$, an issue that will be further investigated in the next subsection.

\begin{figure}[ht]
\caption{Correlogram of National Temperature $X$, $\hat \lo_X$ and $\hat \hi_X$}
\label{fig:acf}
\begin{center}
    \includegraphics[width=12cm,height=8cm]{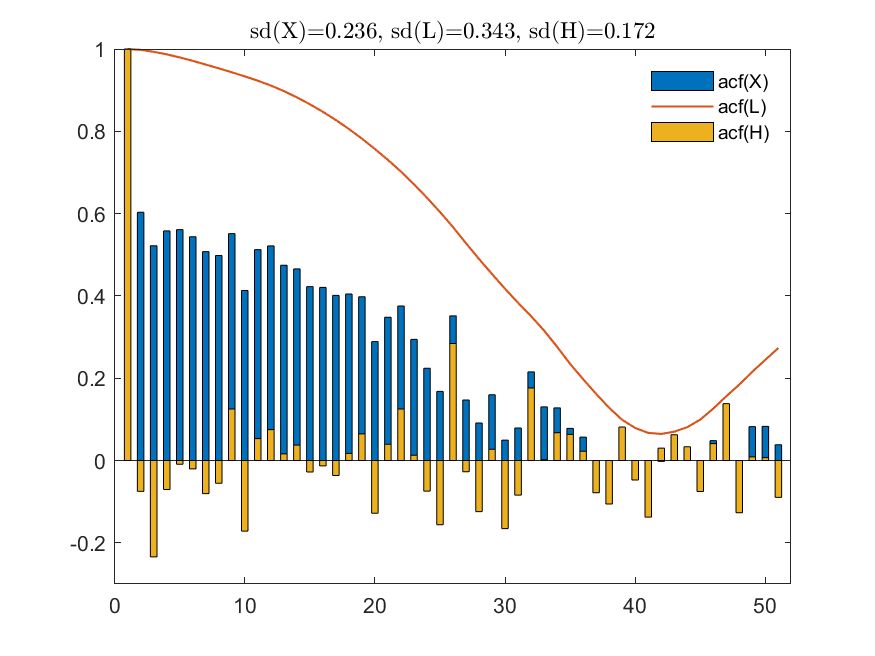}
\end{center}
\footnotesize{$\hat \lo_X$ series is estimated from the UC model with $\sigma_\lo=0.1$, and $\hat \hi_X=X-\hat \lo_X$. With $T=129$,  the standard error under the null hypothesis of no autocorrelation is $\frac{1}{\sqrt{T}}=0.088$.}
\end{figure}

Recognizing spatial differences in the concentration of the two components is  important due to their vastly different dynamic properties. To illustrate this point, Figure \ref{fig:acf} plots the correlogram for $X$ (in blue), $\hat \lo_X$ (in red), and the high-frequency component $\hat \hi_X$ (in brown), at the national level for the United States. The estimate of the low-frequency component $\hat \lo_X$ is based on the UC model with $\sigma_\lo=0.1$. The autocorrelation function of  $\hat \lo_X$  decays very slowly and declines to zero after 35 years. We also notice that $X$ is less persistent than $\hat \lo_X$, which follows from the fact that $\hat \hi_X$ – which exhibits short memory – is included in $X$. Since shocks to $\hat \lo_X$ will last much longer than shocks to $\hat \hi_X$, this motivates us to estimate the economic effects of $\lo^0_X$ and $\hi^0_X$ separately. 

A stark conclusion of our parametric analysis is that $\hat \lo_X$ is strongly persistent and decisively non-stationary, which means it will not mean-revert after an increase in $\epsilon_{\lo}$. In contrast,  $\hat\hi_X$ is nearly white noise and will quickly return to its mean after an increase in $\epsilon_\hi$. Thus, if $\hat\lo_X$ is found to co-vary with economic outcomes, changes in $\hat\lo_X$ will have long-lasting effects. Before proceeding to such an analysis, we need to verify that the strong persistence of $\hat\lo_X$ is not specific to the estimates of the UC model.

 \subsection{Alternative Decompositions}
 Given  the near observational equivalence of many configurations of $(d,\nu)$,  pinning down the `best' parametric model is futile. Nevertheless, we can  estimate the impact of $\lo^0_X$   on the economy without knowing its precise structure. In this subsection,  we consider  trend-filtering procedures used in economic analysis.
 Each procedure requires tuning parameters, so we may think of $\lo_X=\lo_X^0+a_\lo$, where $a_\lo$ is an approximation error.   In practice, $\lo_X$ has to be estimated, and $\hat \lo_X$ will have both an approximation and a sampling error.
 
 We consider three procedures: JH(p,h), HP($\lambda$) or bHP, and MWq. The first is the JH(p,h) procedure of \citet{hamilton:18}  which does not require knowledge of whether $d$ is 0, 1, or 2. The second is the HP($\lambda$) filter of \citet{hodrick-prescott}, and  an improved bHP($\lambda$) version developed  by \citet{phillips-jin:21} and \citet{phillips-shi:21}.    The third is the MWq procedure proposed by \citet{mueller-watson-ecma:08,mueller-watson:17,mueller-watson:22}. MWq   generates a low-frequency component of an arbitrary time series $Z$ with periodicity longer than $2T/q$ by   projecting   $Z$ on a constant and  $q$ periodic functions  collected into $\Psi_T=(\Psi_1,\ldots,\Psi_q)'$, where  $\psi_j(s)=\sqrt{2}\cos(j s\pi)$ for $j=1,\ldots,q,  s=(t-1/2)/T$, for $t=1,\ldots T$.    \citet{mueller-watson-restud:16} show that  the  procedure is valid for  $(b,c,d)$ processes with spectrum $S(\omega)\propto (\omega^2+c^2)^{-d}+b^2$ for $-1/2<d<3/2$. It  approximates $\lo^0_X$ without knowledge of $b,c,$ or $d$ and nests the local-to-unity model and the local-level model as special cases.  

\begin{figure*}[t!]
    %\centering
    \caption{Estimates of $\lo_X$ using MW, HP, and JH.}
    \label{fig:sbs_common_annual}
    \begin{subfigure}[t]{.5\textwidth}
        \caption{$\hat{\lo}$ Aggregated over States}
        \centering
        \includegraphics[width=1\textwidth]{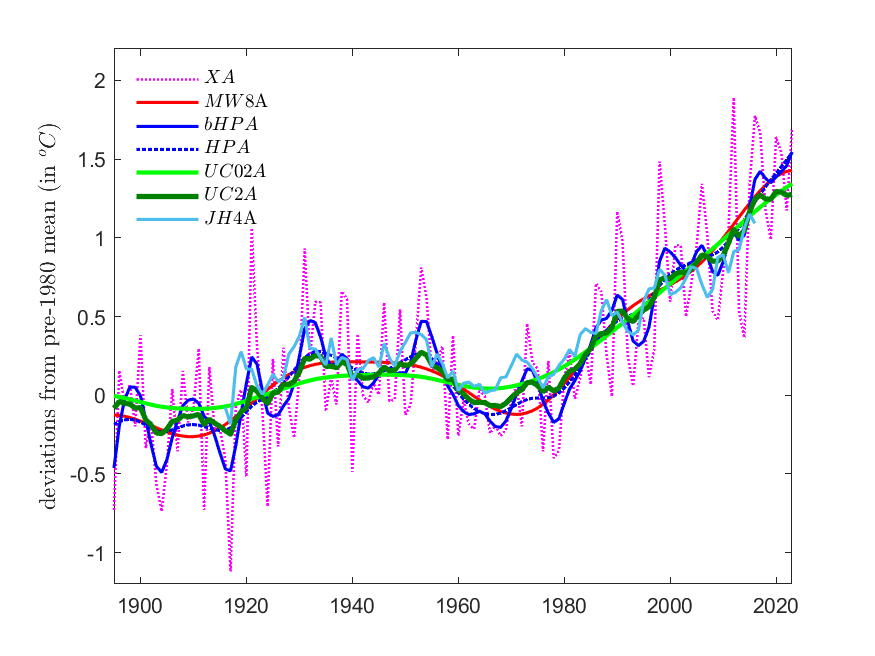}
    \end{subfigure}%
    \hfill
    \begin{subfigure}[t]{.5\textwidth}
        \caption{Common Factors $\hat F$ and National $\hat{\lo}$}
        \centering
        \includegraphics[width=1\textwidth]{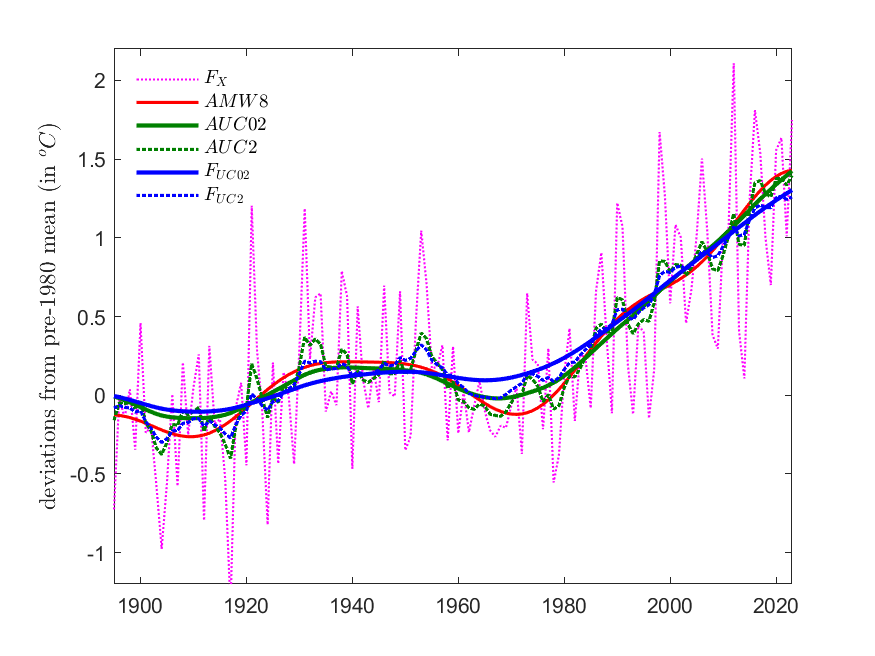}
    \end{subfigure}
\footnotesize{The figure presents U.S. aggregate temperature data XA (as deviation of temperature (in Celsius) from its pre-1980 mean) and various estimates of the low-frequency component $\lo_X$, described in the text.}
\end{figure*}

We will refer to two different but related estimates of aggregate $\hat \lo_X$. When we form aggregate estimates from a weighted sum of state-level estimates $\hat \lo_X = \sum_{i=1}^N \omega_i \hat \lo_{X,i}$ (where $\omega_i$ are population weights), these estimates will be denoted with a label that ends with `A'. When we form an aggregate (or national) series $AX = \sum_{i=1}^N \omega_i X_i$, and then apply an estimation method, the label will start with `A'. For example, AMW8 is obtained by applying the MW procedure with $q=8$ to the aggregate series, while MW8A is obtained by aggregating the 48 state-level  estimates based on MW8. Similarly, UC02A is obtained by aggregating the UC estimates with $\sigma_\lo$ fixed at 0.02, while AUC02 is obtained by applying the estimation method to the aggregate series with the same $\sigma_\lo$. Finally, `$F_Z$' will denote the common factor estimated as the first principal component of a generic series $Z$.

The left panel of Figure \ref{fig:sbs_common_annual}  presents the aggregate of state-level estimates while the right panel of Figure \ref{fig:sbs_common_annual} plots the common factors or the estimates from the national series. In brief,  the MW8A,  HPA(100), and UC02A exhibit similar smoothness, while bHPA, UC2A, and JH4A, are  more variable. However, all estimates trend upward over the course of the sample. A detailed analysis of the properties of these estimates is provided in \ref{app:AppendixL}.

 The procedures considered are designed for filtering economic data. We do not know how accurate they are for estimating temperature trends, and for this, we need to approximate the true  low-frequency component, which is never observed. We thus design  a  Monte-Carlo exercise, calibrated to several temperature series, to evaluate the total (approximation plus sampling) mean-squared error of the procedures.
The simulation exercise, also detailed in \ref{app:AppendixL},  reveals that though the tuning parameters used to obtain economic trends or cycles  are not appropriate for estimating the low- and high- frequency components of temperature data,  any of the methods can produce a similar trend with the right choice of tuning parameter. The issue, then, is to decide on a robust choice of tuning parameters. 

We use the MW procedure for several reasons. First and foremost is that the estimates of $\lo_X$ are quite insensitive to the choice of $q$. Table \ref{tbl:montecarlo} in \ref{app:AppendixL} shows that MW8 (MW with $q=8$) provides a very good approximation of the low-frequency component across different specifications (values of $\sigma_\lo$) and states.
%\st{Furthermore, the same $q$ can be used to generate a similar $\hat\lo_X$ whether we work with monthly, quarterly, or annual data.} 
In addition, the procedure is invariant to aggregation, meaning that AMW8 is the same as MW8A. Lastly, the MW trend is easy to interpret, as we only need to map $q$ to a certain periodicity to understand which atmospheric cycles of interest are captured in the estimate. But  the MW  approximates $\lo_X$ using deterministic functions, and is thus not suited for tracing out its dynamic response to shocks. For such exercise, the stochastic components estimated by UC are more useful.

\begin{figure*}[t!]
    \caption{U.S. Heatmap of $\sigma(X)$ and $\sigma(\hat\lo_X)$ (1895--2023)}
    \label{fig:us_tmp}
    \begin{subfigure}[t]{\textwidth}
        \centering
        \includegraphics[width=0.8\textwidth,trim={0cm 4.5cm 0cm 4.5cm},clip]{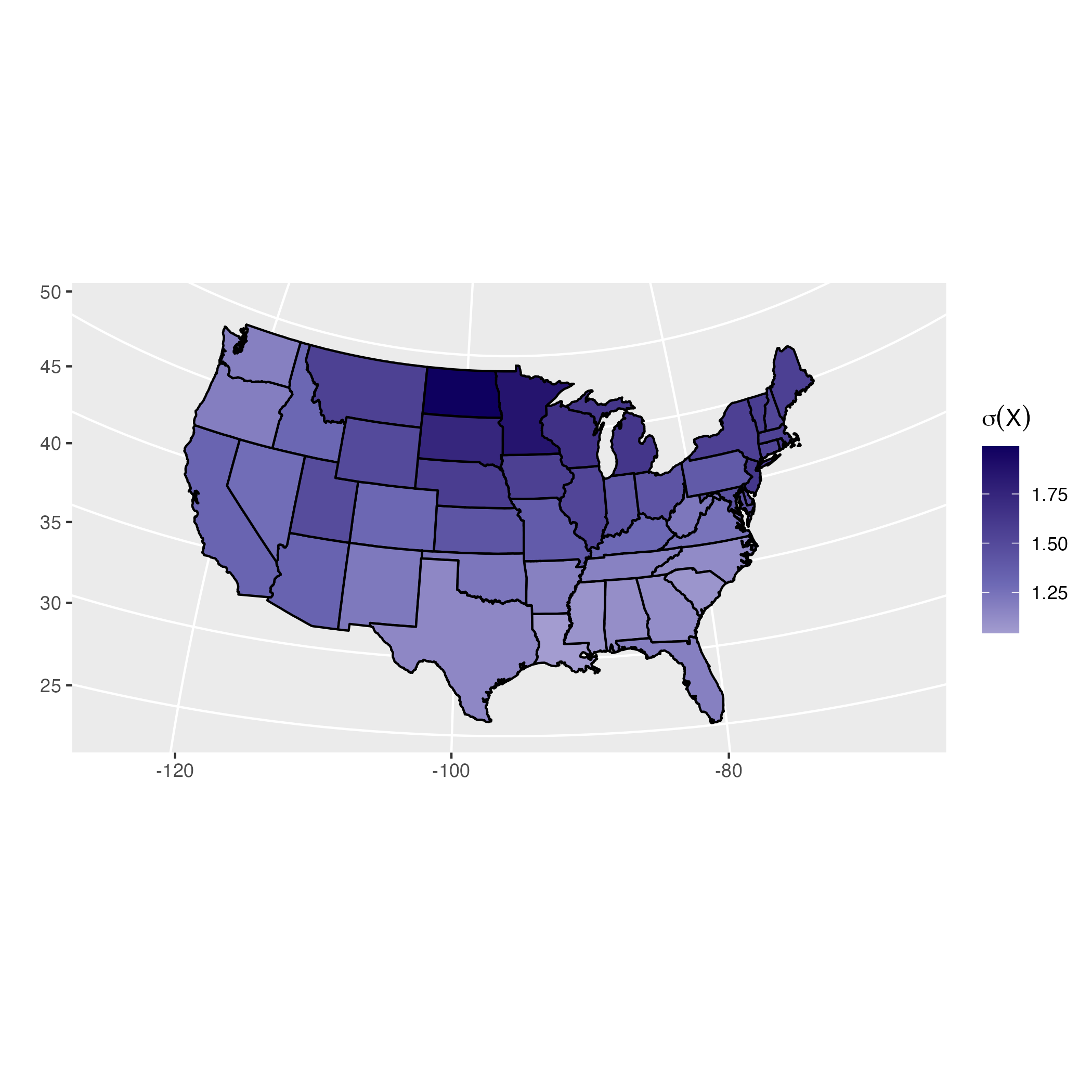}
    \end{subfigure}
    \begin{subfigure}[t]{\textwidth}
        \centering
        \includegraphics[width=0.8\textwidth,trim={0cm 4.5cm 0cm 4.5cm},clip]{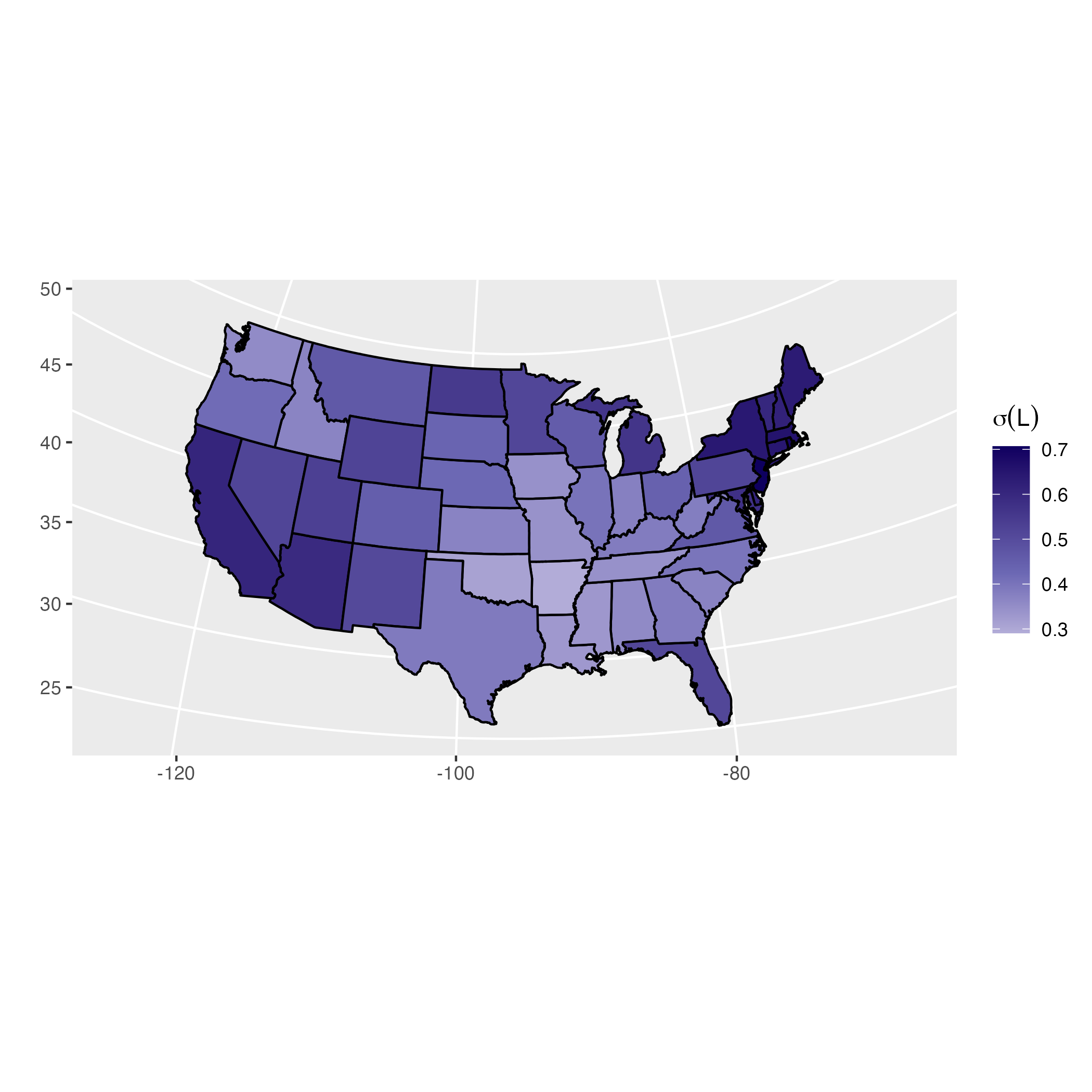}
    \end{subfigure}
    \footnotesize{The top panel plots the standard deviations of  temperature for 48 states in the U.S. The bottom panel plots the standard deviations of the corresponding low-frequency components estimated by MWq with $q=8$.}
\end{figure*}

To be sure that the MWq estimates  convey the same insights as the parameteric model and to appreciate why it is useful to separate temperature data $X$ into two components, consider again annual  temperature for the 48 contiguous states in the U.S. The top panel  of Figure \ref{fig:us_tmp} shows the standard deviation of $X$ for each state over the sample 1895-2023. At face value, the plot suggests that variability  in temperature is more pronounced in the middle of the country. Now consider the bottom panel of Figure \ref{fig:us_tmp} which shows the standard deviation of $\hat \lo_{X}$, where $\hat\lo_X$ is based on the MWq for $q=8$. With $T=129$,  $\hat \lo_X$ captures variations with  periodicity longer than $2T/q\approx 32$ years. Unlike the top panel, this figure indicates that low-frequency variations  in the U.S. are stronger  along the  coasts,  while the high-frequency variations are stronger in the interior. As well, the periodicity of 32 years is also consistent with the shape of the autocorrelation function in  Figure \ref{fig:acf}. 
As we will see below, spatial differences in the concentration of the two components are also found in an international panel spanning 1952-2018.

 The MWq procedure is also useful to learn more about   $\hat \hi_X$.  Because $\hat \hi_X$ is too volatile to be visually informative, we  apply the MWq procedure with a large $q$ in search of insights. With $T=129$, a $q$ of 56 and 26 would correspond to a periodicity of 4.60 and  7.16 years, respectively.
The national estimate of $\hi_X$ is roughly sandwiched by AMW26 and AMW56.   Given that  the ENSO (El Niño--Southern Oscillation) effects are irregular cycles that recur every 2 to 7 years,\footnote{Data source: \url{https://psl.noaa.gov/enso/enso_101.html}}  the  data suggest an ENSO component in $\hat \hi_X$.

\subsection{Common and Idiosyncratic Low-Frequency Variations}
The finding that most states have a trending $\hat \lo_{X,i}$  raises the question of whether these variations are common or state-specific.   We construct the first principal component, denoted $F_X$,  $F_\lo$, and $F_\hi$, from $X$,  $\hat\lo_X$, and $\hat \hi_X$, respectively. The first two are shown in the right panel of Figure \ref{fig:sbs_common_annual}. 

Three features are of note. First, while  $F_X$ in this data  is indistinguishable from the aggregate data $XA$ (equivalently, $AX$),  it is noticeably different  from $F_\lo$, which is smoother.
%Whereas $F_X$   accounts for about 65\% of the variation in  $X$,   $F_\lo$    explains    78\% (JH and bHP) to 90\% (MW8 and HP(100)) of $\hat \lo_X$. . 
Second, estimates of $\lo_X$ using $AX$ are similar to aggregated sub-national estimates, as seen from comparing AUC02 with  UC02A, and AMW8 with  MW8A. Therefore, if one is solely interested in an economy-wide estimate of $\lo^0_X$, it suffices to go top-down and estimate it directly from national data.
% %\footnote{These series are shown in  Figure \ref{fig:sbs_common_fhat} in \ref{app:AppendixD}.}.    
 Third, regardless of the method used, the aggregate estimates  of $\hat\lo_X$ are relatively flat between 1895 and late 1970s, and trend up thereafter.  This is important because we will be analyzing the relationship between  economic outcomes (growth rate of gross state product (GSP) per capita, $\Delta Y$) and $\hat\lo_X$ in the period 1964-2023, where the series $\hat\lo_X$ exhibits strong trending behavior.

\citet{fhsw-jpe:22}  show that low-frequency data admit a  factor representation:\footnote{The  $X$ in \citet{fhsw-jpe:22}  is  labor input and $\Delta Y$ is sectoral TFP.}
\[
\begin{pmatrix} \mathbb X_{i} \\ \Delta  \mathbb Y_{i} \end{pmatrix}=
\begin{pmatrix}
\Lambda_{\mathbb Xi}'I_q & 0 \\ 0 & \Lambda_{\mathbb Yi}'I_q \end{pmatrix} \begin{pmatrix} \mathbb F_1 \\ \mathbb F_2 \end{pmatrix} + \begin{pmatrix} U_X \\ U_Y\end{pmatrix}.
\]
We estimate $\mathbb F$ and the $\Lambda$ matrices  using  the Bayesian procedure outlined in \citet{mueller-watson:22}. The Bayesian estimates of $\mathbb F$ are similar to the principal component estimates shown earlier. However,
 variations in $\lo_X$ over time are due entirely to the $q<<T$ vector $\mathbb X=T^{-1}\Psi_T' X$, and similarly for $\Delta Y$. Because the GSP growth data are only available since 1964, $T=60$ observations may not be very informative about swings with periodicity of 32 years that we used earlier with the longer time span of the temperature data. For this reason, we use $q=8$ for both variables which implies periodicity of 15 years. But information is quite limited even for variations implied by $q=8$. Bayesian inference more adequately reflects this small-sample problem.

\begin{figure}[t!]
\caption{Estimates of Covariability and Factor Loadings by State}
\label{fig:fhsw_lrcov}

    \begin{subfigure}[b]{\textwidth}
        \centering
        \caption{$\hat{\Lambda}_{\mathbb{X}i}$}
        \includegraphics[width=.95\textwidth]{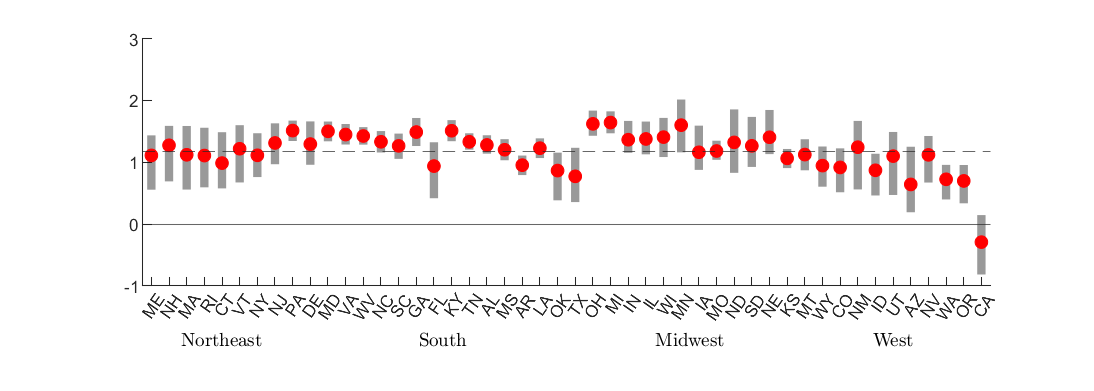}
    \end{subfigure}% 
    \vfill
    \begin{subfigure}[b]{\textwidth}
        \centering
        \caption{$\hat{\Lambda}_{\mathbb{Y}i}$}
        \includegraphics[width=.95\textwidth]{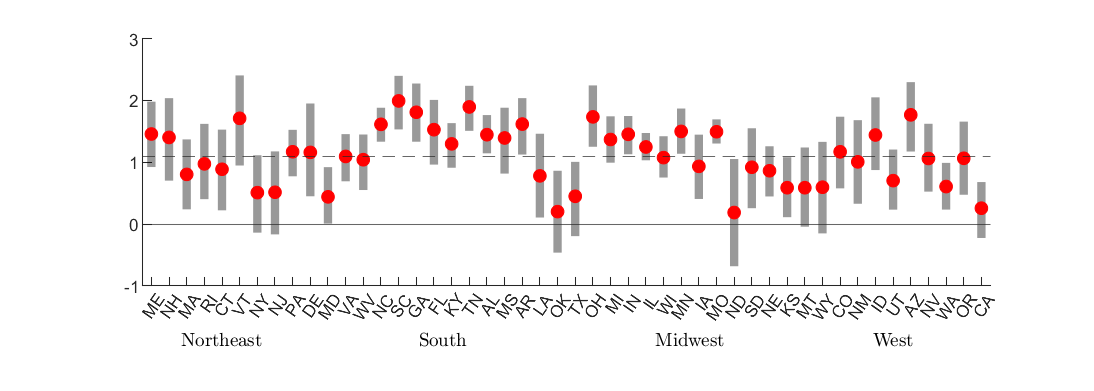}
    \end{subfigure}% 
   
    \vfill

\footnotesize{This figure presents estimates (red dots) by state, corresponding 68\% credible sets (shaded gray bars) and cross-sectional average (dashed line). The sample period is 1964-2023. The top chart (a) presents the estimated loadings associated with temperature and the bottom chart (b) presents the estimated loadings associated with the GSP (gross state product) growth rate. The U.S. states are ordered by region.}
\end{figure}

Figure \ref{fig:fhsw_lrcov} plots the values of $\hat\Lambda_{\mathbb X}$  and $\hat\Lambda_{\mathbb Y}$. To facilitate visualization of spatial patterns, we ordered the states (horizontal axis) by region: Northeast, South, Midwest and West. The 68\% credible regions shaded in grey show that  factor loadings $\hat \Lambda_{\mathbb X}$ for temperature are more precisely estimated than the $\hat \Lambda_{\mathbb Y}$ for GSP growth. Averaged across states, $\hat\Lambda_{\mathbb X}$ and $\hat\Lambda_{\mathbb Y}$ are both around one. However,  $\hat\Lambda_{\mathbb X}$ has a cross-sectional standard deviation of 0.226, much smaller than the 0.472 obtained for $\hat\Lambda_{\mathbb Y}$.  It is interesting that  $\hat \Lambda_{\mathbb X}$ is large  for the states along the coast\footnote{Relatedly, using county-level data for the U.S., \cite{bilal-rossi-hansberg:23} document a four-fold increase in the probability of storms in coastal counties from the 1°C global warming in the last century.}  and  smaller for states inland, while $\hat \Lambda_{\mathbb Y}$ is  larger in the East Coast than the West Coast. This observation is corroborated by a heatmap of  $R^2$ from a regression of  $\mathbb X_i$ on its common component $\hat \Lambda_{\mathbb Xi}' \hat F$ (not shown), 
%in Figure \ref{fig:fhsw_heatmap} in the Appendix, 
suggesting that  idiosyncratic variations dominate the common variations for states inland.  Also, consistent with  Figure \ref{fig:us_tmp}, states that have large changes in the low-frequency component over time  are also the ones with a dominating common component. The results also suggest substantial heterogeneity across states. 

\begin{figure*}[ht!]
\caption{Aggregate Variations by Components for U.S. Data}
    \label{fig:fhsw5_us}
    \begin{subfigure}[b]{0.5\textwidth}
        \centering
        \caption{GSP Growth}
        \includegraphics[width=1.1\textwidth]{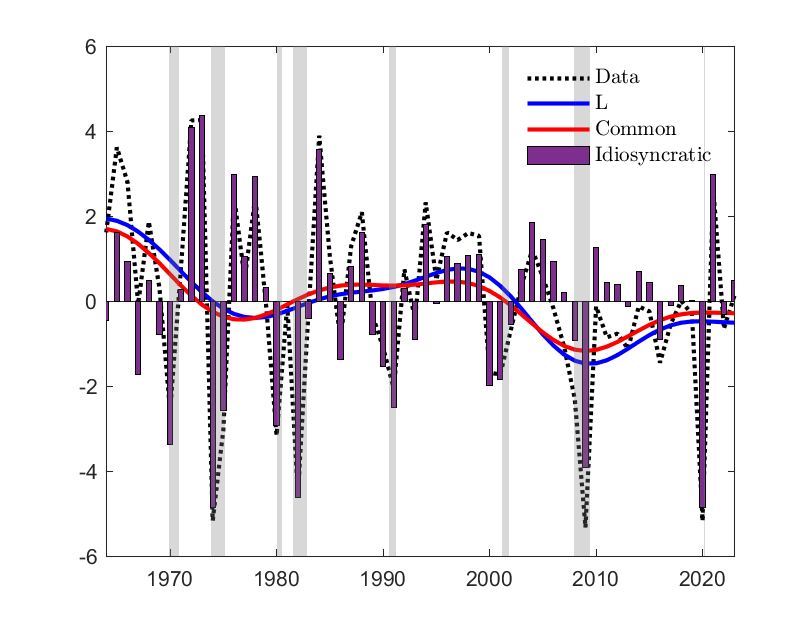}
    \end{subfigure}%
    ~ 
    \begin{subfigure}[b]{0.5\textwidth}
        \centering
        \caption{Temperature}
        \includegraphics[width=1.1\textwidth]{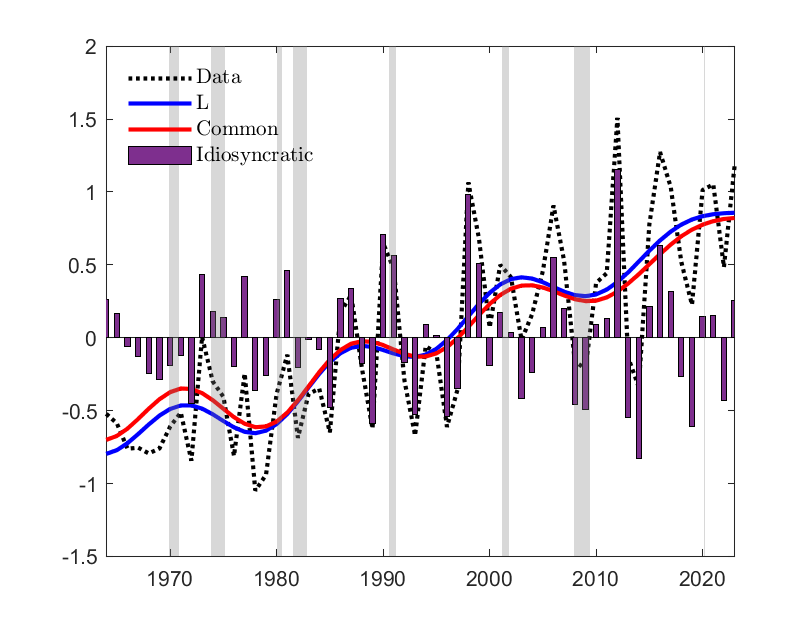}
    \end{subfigure}
        \vfill
\footnotesize{This figure presents the actual series (black dashed line), its low-frequency component estimated using MW8 (in blue),  as well as its decomposition into  a common component (in red) and idiosyncratic error (in brown bars) using the framework of \citet{fhsw-jpe:22}).  The two variables are aggregate GSP growth $\Delta Y$ (left) and aggregate temperature $X$ (right). The sample period is 1964-2023. Shaded areas denote NBER recessions.}
\end{figure*}
Figure \ref{fig:fhsw5_us}  presents a common-idiosyncratic view of  the aggregate data. The left panel  plots the deviation of aggregate GSP growth from its mean (in black dots), the low-frequency component of aggregate GSP growth estimated by MW8 (in blue), and a decomposition of the low-frequency component into a  common (in red) and  idiosyncratic (in brown bars) component using the framework of \citet{fhsw-jpe:22}. Evidently, the idiosyncratic component of growth aligns with recessions in 1975, 1982, 2008, and 2020, while the low-frequency component moves closely with the common component. The right panel of Figure \ref{fig:fhsw5_us} plots the same components for the national temperature series. Many of the spikes in the temperature data coincide with  the idiosyncratic component, while the  low-frequency component aligns with the common component.  The eye-catching feature in  Figure \ref{fig:fhsw5_us}  is   the wavy downward trend in $\hat \lo_{AY}$ (left panel) and  the stark upward trend in $\hat \lo_{AX}$ (right panel).\footnote{It is also worth noting that the Durbin-Watson statistic from regressions of the log level of aggregate GSP on temperature is small, suggesting that the relation might be spurious.}  The next section considers if this apparent negative relationship is statistically significant using regressions.

\section{Estimating the Economic Impact of Warming Temperature}
\label{sec:Sect4}
 
There is a large body of work studying the impact of environmental changes on the economy.\footnote{See \citet{dell-jel:14} for a review of the methodologies used and \citet{tol-ep:24} for a review of the estimates.} Panel regressions of $\Delta Y$ on  $\Delta X$ are often interpreted as short-run estimates, while  regressions of the long-difference of $Y$ on the long-difference of $X$ are taken as longer-run estimates. However, the notion of long- and short-run is not made precise.  \citet{hsiang-burke:14} drew attention to a ``frequency-identification tradeoff.'' The issue is that  low-frequency  data can account for  adaptation, but the assumption of a constant economic structure is not easy to defend. In contrast, high-frequency data can more credibly make this assumption, but cannot capture the effects of adaptation which takes time. 
 
%\citet{tol-ep:24} \st{reviewed the  estimates and finds that while the impact of warming is generally  negative, there are considerable differences in the magnitude  between those with $X$ and those with $\Delta X$ as regressors. This is to be expected as zero-frequency variations are removed from the latter, while the former maintains a strong low-frequency component. Second, most econometric analyses control for individual or year effects in an additive way.}

 We also take the view that variation of $X$ at different frequencies   can have different impact on economic outcomes, and our approach is to use both $\hat \lo_X$ and $\hat \hi_X$ as regressors while keeping $\Delta Y$, which is comprised of variations from all frequencies, as a dependent variable.  \citet{newell-prest-sexton:21},  \citet{berg-curtis-mark:24}  and \citet{kahn-etal:23} also use GDP growth as an outcome variable but few studies entered the time variation of both components in the same regression. \citet{bento-etal:23} enter both the mean and deviations as regressors, but using the full sample mean as $\hat\lo_X$ which varies across $i$ but not $t$. \citet{kahn-etal:21} do not include the (trending) rolling mean as regressor due to  spurious regression concerns.    
  \citet{berg-curtis-mark:24}  assume that $\lo^X_0$ is exogenous so that the estimates remain asymptotically normal in spite of being non-stationary. Concerned about the use of year effects in these regressions, \citet{berg-curtis-mark:24} analyze international data using a common-idiosyncratic decomposition of $X$. Our focus is on a decomposition of the frequency components  rather than  common-idiosyncratic decomposition of $X$.

\subsection{The Regression Framework}
Suppose that we have  national  data, or data for a   single country or state. An infeasible  regression to learn about the long-run  effect of $\lo^0_X$ on $\Delta Y$ is 
\[ \Delta Y_t=\beta_0 +\beta_\lo\lo^0_{X,t}+  u^0_{Y,t}.\] 
If the low-frequency component $\lo^0_{X}$ is $I(1)$ and $u^0_Y$ is stationary, the above static regression would consistently estimate the cointegrating relation between the two variables.  We only assume that  $\lo^0_X$ has non-trivial variations at frequency close to zero, but do not require $d$ to be integer valued. The feasible  regression replaces $\lo^0_X$ with an estimate $\hat \lo_X$, 
\begin{equation}
\Delta Y_{t}=\beta _{0}+\beta _{\mathcal{L}} \hat \lo_{X,t}+u_{Y,t}.
\label{eq:coint}
\end{equation}
Note that  $\hi^0_X$ is consolidated into the error term $u^0_Y$ in this model, as its  variations are, asymptotically, of lower order than $ \lo^0_{X}$. An appeal of the MW estimates $\hat\lo_X$ and $\hat \hi_X$ is that they are orthogonal by construction; including $\hat \hi_{X}$ in (\ref{eq:coint}) does not affect the estimate of $\beta_\lo$, but can improve efficiency.

The  model  for a single unit  motivates our panel regressions of the form
\begin{eqnarray}
\Delta Y_{it}&=&m _{it}+\beta _{\mathcal{L}}\mathcal{\hat{L}}_{X,it}+u_{Y,it}.
\label{eq:static}
\end{eqnarray}
As in \citet{berg-curtis-mark:24}, we assume that $\lo^0_X$ and $\hi^0_X$ are exogenous. Exogeneity of the high-frequency component seems natural. While exogeneity of the low-frequency component may seem concerning in view of the issue raised in \citet{hsiang-burke:14},   $\lo^0_X$ is non-stationary and thus endogeneity bias is of second order. The low-frequency covariability between $X$ and $\Delta Y$  shown earlier also eases  our concern of a spurious regression, and  we guard against  omitted correlations by
considering different controls for unobserved heterogeneity $m_{it}$.

We consider three specifications:
\begin{itemize}
\item[i.] Model FE: \;$m_{it}=\gamma_i;$
\item[ii.] Model AFE: \;$m_{it}=\gamma_i+\xi_t;$
\item[iii.] Model IFE:\; $m_{it}=\gamma_i+\lambda_i F_t.$
\end{itemize}

Model FE  only controls for unit fixed effects, which would be appropriate if there were no common year effects. Comparing these estimates with model AFE, which also controls for year effects, can help  gauge the importance of $\xi_t$. Applied work often uses one-way clustered standard errors for inference, where clustering is only performed at the unit level. However, as discussed in \cite{majerovitz-sastry:23}, one-way clustering  may not sufficiently characterize all sources of heterogeneity, in particular in the time dimension.  Our simulation results in \ref{app:AppendixB} illustrate this concern. Specifically, we find that if the common year effect is omitted or affects the states in a non-uniform way, one-way clustered standard errors tend to understate the estimation uncertainty and lead to severe over-rejections of the null hypothesis (up to 80\% at 10\% nominal levels). In such cases, valid inference in FE and AFE requires two-way clustering in order to account for strong time effects that do not affect the states uniformly. We use the two-way clustered standard errors based on \citet{CGM2011}.

An alternative to correcting the standard errors for correlations in the units temporally and cross-sectionally is to  model the correlation directly using an interactive fixed effects (IFE) specification. In this specification, $\lambda _{i}$ is an $r$-vector of factor loadings and $F_{t}$ is an $r
$-vector of common factors. In our analysis, we set $r=1$. The estimation is
performed using the iterative procedure suggested by \citet{bai:09}.  In this model, the time effects are no longer a simple cross-sectional average at each $t$ (as in model AFE), but are determined by principal component analysis. To account for the relatively small sample used to estimate the latent factor structure, we use a fixed-design, percentile wild bootstrap method that preserves the complex dependence structure in the data. For the IFE estimates, we report the  90\% and 68\%
bootstrap-based confidence intervals, which account for the uncertainty from estimating the latent factor structure.

The parameter  $ \beta_\lo$  in the panel regression given by (\ref{eq:static}) measures the average long-run impact of $\hat \lo_{X,it}$ (see  \citet{phillips-moon:99}). To estimate the short-run effects, we use an autoregressive distributed lag  model, with the same specifications for $m_{it}$ noted above,
\begin{eqnarray}
\Delta Y_{it}&=&m _{it}+\alpha \Delta Y_{i,t-1}+b_{\mathcal{L}}\mathcal{\hat{L}}%
_{X,it}+\delta_{\mathcal{H}}\Delta \hat{\mathcal{H}}_{X,it}+\gamma_{\hi}\mathcal{\hat{H}}_{X,i,t-1}+u_{Y,it}.
\label{eq:dynamic}
\end{eqnarray}
The parameters $\gamma_{\hi}$ and $\delta_\hi$   measure the sum  and impact of the  high-frequency  components on economic growth, respectively. By including $\Delta Y_{i,t-1}$, the dynamic regression not only  controls for feedback, but also guards against the  bias discussed in  \cite{klosin:24}.\footnote{Because $\Delta \hat \lo_{X,it}$ is small in this data, $\hat \lo_{X,it-1}$ is highly correlated with $\hat \lo_{X,it}$, and is therefore omitted from the regression.}  The quantity $\frac{\hat b_{\lo}}{1-\hat\alpha}$ estimates a long-run effect of a permanent change in $\hat \lo_{X}$ but inference based on the ratio of two asymptotically normal estimates can be inaccurate.\footnote{See, for example, \citet{reed-zhu:17} for a discussion.} The static model given in (\ref{eq:static}) provides a direct but less efficient estimate of the long-run effect. When complemented with an equation that specifies the dynamics of $\hat \lo_{X,i}$,  equation (\ref{eq:dynamic}) can be used to trace out the effect of a change in $\lo_{X,i}$ at any horizon. We will consider such an exercise below.

\subsection{Estimates for U.S. State Panel}

In this subsection, we present empirical results 
%on the economic effects of the
%low- and high-frequency movements in temperature on economic activity using
based on a panel of U.S. state level data for the 1964--2023 sample. The U.S. state panel has been used in other studies. \citet{bilal-rossi-hansberg:23} developed a spatial dynamic model to explain the aggregate and local cost of natural disasters,  which are extreme high frequency variations.     \citet{kahn-etal:23} estimates the output effects of deviations from rolling means,  while \citet{colacito-hoffman-phan:19} consider seasonal temperature changes.   Our focus on low and high frequency variations is distinct. 
To facilitate comparisons with the international panel, $X_{it}$ and its components are converted in
Celsius while $\Delta Y_{it}$,
is a growth rate in percent per year. The low-frequency components of both variables are estimated by the MW method with a tuning parameter $q=8$. Given that $T$ for this sample is 60, $\hat \lo_X$ now has a periodicity of around 15 years. The regression results are, however, quite robust to the choice of $q$.\footnote{
Alternatively, we could
estimate the two latent components using the conventional state-space model or the \citet{hartl-uc} method. The results based on these alternative estimates, as well as those based on the MWq estimates with $q=4$ and 12,
are qualitatively similar and are available upon request.}
Table \ref{tbl:USpanel} presents results for the parameters of interest: $\beta_\lo$ from the static specification, and  $b_\lo$, $\delta_\hi$, $\gamma_\hi$ and $\alpha$ for the dynamic specification.  

\begin{table}
\caption{U.S. Panel Data Regressions}
\label{tbl:USpanel}
\begin{center}
\begin{tabular}{l|c|cccc}
& static & \multicolumn{4}{|c}{dynamic} \\ \hline\hline
& $\beta _{\mathcal{L}}$ & $b_{\mathcal{L}}$ & $\delta _{\mathcal{H}}$ & $\gamma_{%
\mathcal{H}}$ & $\alpha $ \\ \hline\hline
FE &  &  &  &  &  \\ 
estimate & $-0.868$ & $-0.753$ & $-0.105$ & $-0.127$ & $0.123$ \\ 
1way SE & $(0.092)$ & $(0.095)$ & $(0.073)$ & $(0.096)$ & $(0.029)$ \\ 
2way SE & $(0.544)$ & $(0.522)$ & $(0.268)$ & $(0.455)$ & $(0.088)$ \\ 
90\% CI & $[-1.770,0.029]$ & $[-1.617,0.090]$ & $[-0.576,0.351]$ & $%
[-0.879,0.636]$ & $[-0.028,0.270]$ \\ 
68\% CI & $[-1.434,-0.361]$ & $[-1.311,-0.242]$ & $[-0.392,0.178]$ & $%
[-0.608,0.338]$ & $[0.038,0.212]$ \\ \hline\hline
AFE &  &  &  &  &  \\ 
estimate & $-0.126$ & $-0.112$ & $-0.104$ & $-0.007$ & $0.129$ \\ 
1way SE & $(0.287)$ & $(0.259)$ & $(0.085)$ & $(0.158)$ & $(0.032)$ \\ 
2way SE & $(0.425)$ & $(0.397)$ & $(0.134)$ & $(0.257)$ & $(0.045)$ \\ 
90\% CI & $[-0.833,0.580]$ & $[-0.798,0.569]$ & $[-0.356,0.148]$ & $%
[-0.415,0.421]$ & $[0.057,0.209]$ \\ 
68\% CI & $[-0.579,0.289]$ & $[-0.546,0.311]$ & $[-0.257,0.048]$ & $%
[-0.256,0.249]$ & $[0.084,0.174]$ \\ \hline\hline
IFE &  &  &  &  &  \\ 
estimate & $-0.343$ & $-0.298$ & $-0.217$ & $-0.119$ & $0.164$ \\ 
90\% CI & $[-0.972,0.238]$ & $[-0.952,0.302]$ & $[-0.478,0.060]$ & $%
[-0.542,0.286]$ & $[0.102,0.223]$ \\ 
68\% CI & $[-0.713,0.024]$ & $[-0.680,0.067]$ & $[-0.370,-0.057]$ & $%
[-0.355,0.135]$ & $[0.128,0.200]$ \\ \hline
\end{tabular}

\end{center}
\footnotesize{The table
reports panel regression results for the model with individual fixed effects only (FE, top panel), additive fixed effects with both individual and time effects (AFE, middle panel) and interactive fixed effects with country fixed effects (IFE, bottom panel). Results are presented for the static model (left) and the dynamic specification from Section 3.1 (right). For the FE and AFE models, the table presents point estimates, one-way cluster-robust standard errors (1way SE), two-way cluster-robust standard errors (2way SE), and 90\% and 68\% bootstrap confidence intervals. For the IFE model, the table presents the estimates, and 68\% and 90\% bootstrap confidence intervals.} \medskip

\end{table}

For model FE,  $\beta_\lo$ and $b_\lo$ are estimated to be 
$-0.868$ and $-0.753$, respectively.  In contrast,   the impact effect associated with $\hi_{X,it}$ is only  $-0.105$. The one-way clustered standard errors (1way SE) – clustered at the state level – are quite small and imply highly significant estimates. The more robust two-way clustered standard errors  (2way SE) are several orders of magnitude larger, to the point of eliminating any statistical significance of $\hat\beta_\lo$ and $\hat b_\lo$ at 10\% significance levels. To more accurately approximate the various sources of uncertainty associated with these estimates, we also construct and report 68\% and 90\% (fixed-design) bootstrap confidence intervals for all estimates. The bootstrap confidence intervals lend some support to the possibility that the estimated negative effect of $\lo_{X,it}$ on $\Delta Y_{it}$ is statistically significant in model FE.

The estimates of $\beta_\lo$  and $b_\lo$ for model AFE are closer to zero, suggesting that common time variations have been omitted from model FE.  However, model AFE model assumes a homogeneous response to the common year variations, which may be a restrictive assumption.  The bottom panel of Table \ref{tbl:USpanel}  reports  estimates from  model IFE.  Although 
  $\hat\beta_\lo$ and $\hat b_\lo$ in this IFE model are still negative, they are statistically insignificant. Since the  coefficient $\beta_X$ in a regression of $\Delta Y$ on $X$ is a combination of $\beta_\lo$ and $\beta_\hi$,\footnote{As $\cov(X,\Delta Y)=\beta_\lo \var(\lo^0_X)+ \beta_\hi\var(\hi^0_X)$,  the population parameter $\beta_X$  is
 \[\beta_X= \beta_\lo \frac{\var(\lo^0_X)}{\var(X)} + \beta_\hi \frac{\var(\hi^0_X)}{\var(X)}\]
and the least squares estimate $\hat \beta_X$ will be a  weighted sum of $\beta_\lo$ and $\beta_\hi$.}  our results suggest that if we had performed an AFE or IFE regression with $X$ as the regressor,  $\hat \beta_X$ would also be small and insignificant. Indeed, $\hat \beta_X$ from the two regressions are $-0.111$ and  $-0.198$ respectively. Though negative,  the standard errors and confidence intervals all suggest that $\beta_X$ is not statistically different from zero.

  One interesting result that emerges from Table \ref{tbl:USpanel} is that the impact effect of $\hi_{X,it}$ in the IFE model is estimated to be $-0.217$ and marginally significant with a 68\% bootstrap confidence interval of $[-0.370,-0.057]$. Although $-0.217$ may seem small and the effects of the high-frequency component should be short-lived, the estimated economic magnitude can be large because  the high-frequency component $\hi_{X,it}$ is more variable than its low-frequency counterpart $\lo_{X,it}$. Overall, the relationship between temperature and economic activity in the U.S. panel – to the extent there is one – is more likely to be driven by  high-frequency weather variations such as heat waves or natural disasters.

\begin{figure*}[t!]
\caption{Plot of  $\hat{F}\hat{\protect\lambda}^{\prime }$ in 
Model IFE }
\includegraphics[width=17cm,height=8.5cm]{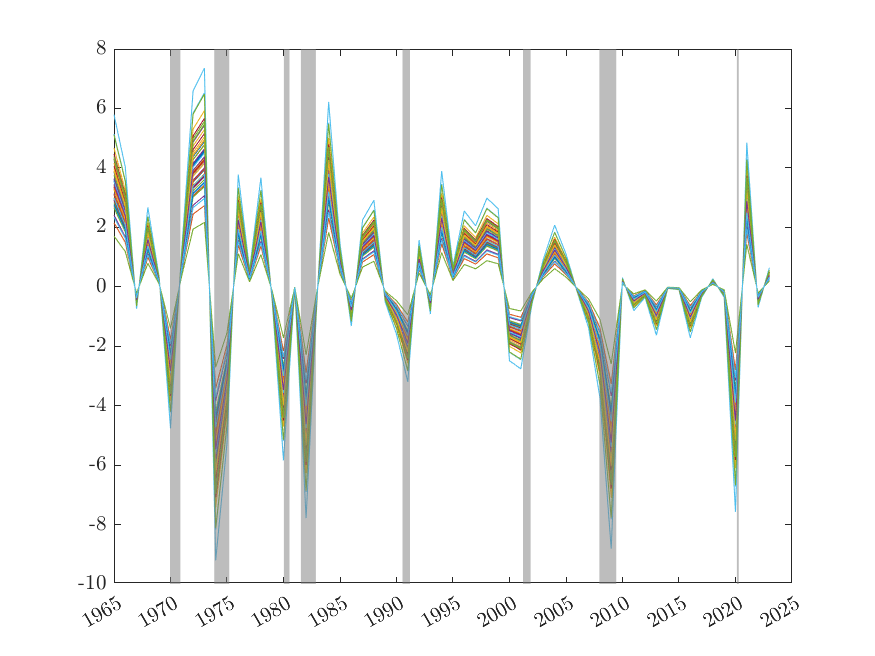}
\label{fig:factstr}

\footnotesize{ This figure plots the interactive fixed effect that is estimated from the IFE model for 48 U.S. states. Shaded areas signify NBER recessions.}
\end{figure*}

To better understand why standard errors are sensitive to the type of clustering, Figure \ref{fig:factstr} plots the estimated interactive fixed effect for all 48 states implied by  model IFE. As the troughs coincide with recession dates, it is clear that $\hat F\hat\lambda'$ is picking up business cycle conditions.  Given that $F_t$ is common at each $t$, the heterogeneity shown in  $\hat\lambda_i$ is inconsistent with the assumption of a common response to $F_t$, underlying model AFE. Thus, the residuals of the additive fixed effect specifications remain correlated and necessitate two-way clustering in the construction of the standard errors. The inclusion of these effects in model IFE helps control for high-frequency variations in GSP growth, making it easier to detect a low-frequency relationship with $\hat \lo_X$.

Finally, to assess the finite-sample properties of the estimators and the inference procedures used in this section, \ref{app:AppendixB} reports Monte Carlo simulation evidence from a model that is calibrated to U.S. data. Despite the relatively small time-series and cross-sectional dimensions, the estimators are relatively well centered around their true values and the bootstrap confidence intervals have reliable coverage rates. 

\subsection{ International Panels}

We further consider an international panel of real GDP growth and temperature as described in \ref{app:AppendixC}. The quality of economic data again drives our decision to select only the top 50 countries from the Data Quality Ratings produced by World Economics. In addition to aligning the quality of the economic data, these countries span all continents, which better captures global climate changes over the sample. We also take 20 countries from the full panel to form an European panel. The main limitation of these data is that economic growth is not available for the recent past, 2019 to 2024. Temperatures are reported in Celsius, and $X_i$ are defined as deviations from their pre-1980 mean. 

\ref{app:AppendixTF} provides summary statistics about the decomposition of $X_i$ into low- and high-frequency components. Some highlights are as follows (see Table \ref{tbl:intl_d}). In the larger international panel with 50 countries, those in the southern hemisphere such as the Philippines, Singapore, Malaysia, Brazil, and Australia have lower values of $\nu=\frac{\sigma_\hi}{\sigma_\lo}$ than countries further north.  As in the U.S., there are significant spatial variations in the European data. The Nordic countries (e.g., Denmark, Finland, Norway, Sweden) tend to exhibit larger volatility in $X$ than those in Southern Europe (e.g., Italy, Greece, Spain, Portugal). While $\hat \lo_{X,i}$ is quite flat in Ireland, Iceland and Great Britain, there is a notable upward trend in $\hat \lo_{X,i}$ for Austria, Germany, and Italy. More generally, low-frequency variations seem to be stronger in Europe than in the U.S. Performing a common-idiosyncratic decomposition (Figure \ref{fig:fhsw5_europe}), one easily sees  a negative correlation between the low frequency component of growth and temperature, similar to Figure \ref{fig:fhsw5_us}.

\begin{table}
\caption{European Panel Data Regressions}
\label{tbl:EUpanel}
\begin{center}

\begin{tabular}{l|c|cccc}
& static & \multicolumn{4}{|c}{dynamic} \\ \hline\hline
& $\beta _{\mathcal{L}}$ & $b_{\mathcal{L}}$ & $\delta _{\mathcal{H}}$ & $\gamma_{%
\mathcal{H}}$ & $\alpha $ \\ \hline\hline
FE &  &  &  &  &  \\ 
estimate & $-1.431$ & $-0.959$ & $0.018$ & $-0.322$ & $0.323$ \\ 
1way SE & $(0.172)$ & $(0.161)$ & $(0.145)$ & $(0.137)$ & $(0.063)$ \\ 
2way SE & $(0.358)$ & $(0.348)$ & $(0.243)$ & $(0.302)$ & $(0.079)$ \\ 
90\% CI & $[-1.982,-0.870]$ & $[-1.530,-0.403]$ & $[-0.361,0.426]$ & $%
[-0.886,0.187]$ & $[0.189,0.444]$ \\ 
68\% CI & $[-1.763,-1.116]$ & $[-1.313,-0.653]$ & $[-0.218,0.264]$ & $%
[-0.657,-0.011]$ & $[0.242,0.396]$ \\ \hline\hline
AFE &  &  &  &  &  \\ 
estimate & $-0.803$ & $-0.568$ & $0.264$ & $0.069$ & $0.271$ \\ 
1way SE & $(0.789)$ & $(0.591)$ & $(0.293)$ & $(0.278)$ & $(0.065)$ \\ 
2way SE & $(0.845)$ & $(0.661)$ & $(0.339)$ & $(0.366)$ & $(0.068)$ \\ 
90\% CI & $[-1.736,0.142]$ & $[-1.474,0.389]$ & $[-0.189,0.742]$ & $%
[-0.574,0.658]$ & $[0.146,0.382]$ \\ 
68\% CI & $[-1.372,-0.213]$ & $[-1.117,-0.012]$ & $[-0.018,0.539]$ & $%
[-0.314,0.417]$ & $[0.198,0.341]$ \\ \hline\hline
IFE &  &  &  &  &  \\ 
estimate & $-0.904$ & $-0.743$ & $0.141$ & $-0.223$ & $0.344$ \\ 
90\% CI & $[-1.477,-0.331]$ & $[-1.299,-0.174]$ & $[-0.289,0.566]$ & $%
[-0.767,0.349]$ & $[0.259,0.433]$ \\ 
68\% CI & $[-1.262,-0.586]$ & $[-1.051,-0.432]$ & $[-0.124,0.401]$ & $%
[-0.572,0.123]$ & $[0.292,0.394]$ \\ \hline
\end{tabular}

\end{center}
%\footnotesize{See notes to Table \ref{tbl:USpanel}.} 
\footnotesize{The table
reports panel regression results for the model with individual fixed effects only (FE, top panel), additive fixed effects with both individual and time effects (AFE, middle panel) and interactive fixed effects with country fixed effects (IFE, bottom panel). Results are presented for the static model (left) and the dynamic specification from Section 3.1 (right). For the FE and AFE models, the table presents point estimates, one-way cluster-robust standard errors (1way SE), two-way cluster-robust standard errors (2way SE), and 90\% and 68\% bootstrap confidence intervals. For the IFE model, the table presents the estimates, and 68\% and 90\% bootstrap confidence intervals.}
\medskip

\end{table}

Table \ref{tbl:EUpanel}, which presents regression results for the European panel, follows the same format as Table \ref{tbl:USpanel}.  The main finding is that the estimates of $\beta_{\lo}$  and $b_{\lo}$ are much larger in magnitude than the ones for the U.S panel and are highly statistically significant. For example, the estimates of $\beta_{\lo}$  and $b_{\lo}$ in the FE specification are $-1.431$ and $-0.959$, respectively. In the static IFE model, $\hat\lo_X$ is estimated to have an effect of  $-0.904$, with  a 90\% bootstrap confidence interval of $[-1.477, -0.331]$. 
The (population-weighted)  average of $\hat \lo_X$ in Europe rose by 1.48°C between 1980 and 2018, suggesting a loss in European GDP growth of around 1.34 percentage points over this 38-year period.
In the dynamic specification, the point estimate of $b_\lo$ is $-0.743$, with a 90\% bootstrap confidence interval of  $[-1.299, -0.174]$. %\textcolor{red}{$b_L$ is not the long run multiplier though.} 
 The first-order effect of $\hat \hi_X$ is not significant while the estimated coefficient on the lagged dependent variable in the dynamic specification, $\alpha$, suggests a higher persistence of GDP growth in the European panel than the U.S. state panel.

\begin{table}
\caption{International Panel Data Regressions}
\label{tbl:Ipanel}
\begin{center}
\begin{tabular}{l|c|cccc}
& static & \multicolumn{4}{|c}{dynamic} \\ \hline\hline
& $\beta _{\mathcal{L}}$ & $b_{\mathcal{L}}$ & $\delta _{\mathcal{H}}$ & $\gamma_{%
\mathcal{H}}$ & $\alpha $ \\ \hline\hline
FE &  &  &  &  &  \\ 
estimate & $-0.969$ & $-0.614$ & $0.103$ & $-0.261$ & $0.274$ \\ 
1way SE & $(0.245)$ & $(0.185)$ & $(0.144)$ & $(0.120)$ & $(0.039)$ \\ 
2way SE & $(0.383)$ & $(0.347)$ & $(0.234)$ & $(0.308)$ & $(0.059)$ \\ 
90\% CI & $[-1.374,-0.325]$ & $[-1.132,-0.116]$ & $[-0.273,0.484]$ & $%
[-0.841,0.275]$ & $[0.180,0.366]$ \\ 
68\% CI & $[-1.186,-0.593]$ & $[-0.932,-0.331]$ & $[-0.130,0.331]$ & $%
[-0.606,0.061]$ & $[0.215,0.330]$ \\ \hline\hline
AFE &  &  &  &  &  \\ 
estimate & $-0.305$ & $-0.212$ & $0.344$ & $0.240$ & $0.258$ \\ 
1way SE & $(0.552)$ & $(0.415)$ & $(0.154)$ & $(0.167)$ & $(0.039)$ \\ 
2way SE & $(0.586)$ & $(0.450)$ & $(0.201)$ & $(0.230)$ & $(0.058)$ \\ 
90\% CI & $[-0.885,0.310]$ & $[-0.763,0.386]$ & $[0.011,0.670]$ & $%
[-0.203,0.651]$ & $[0.166,0.346]$ \\ 
68\% CI & $[-0.666,0.058]$ & $[-0.561,0.143]$ & $[0.156,0.544]$ & $%
[-0.030,0.482]$ & $[0.202,0.312]$ \\ \hline\hline
IFE &  &  &  &  &  \\ 
estimate & $-1.282$ & $-0.751$ & $0.002$ & $0.209$ & $0.250$ \\ 
90\% CI & $[-1.806,-0.835]$ & $[-1.249,-0.280]$ & $[-0.319,0.303]$ & $%
[-0.232,0.623]$ & $[0.181,0.324]$ \\ 
68\% CI & $[-1.588,-1.000]$ & $[-1.042,-0.477]$ & $[-0.187,0.188]$ & $%
[-0.061,0.480]$ & $[0.207,0.295]$ \\ \hline
\end{tabular}

\end{center}

%\footnotesize{See notes to Table \ref{tbl:USpanel}.} 
\footnotesize{The table
reports panel regression results for the model with individual fixed effects only (FE, top panel), additive fixed effects with both individual and time effects (AFE, middle panel) and interactive fixed effects with country fixed effects (IFE, bottom panel). Results are presented for the static model (left) and the dynamic specification from Section 3.1 (right). For the FE and AFE models, the table presents point estimates, one-way cluster-robust standard errors (1way SE), two-way cluster-robust standard errors (2way SE), and 90\% and 68\% bootstrap confidence intervals. For the IFE model, the table presents the estimates, and 68\% and 90\% bootstrap confidence intervals.}
\medskip

\end{table}

The panel regression results for the larger international panel ($N=50$) are reported in Table \ref{tbl:Ipanel}. Consistent with the European panel, the results point to a fairly large negative and statistically significant effect of the low-frequency component of temperature on economic activity. For the FE model,  $\hat\beta_\lo$ and $\hat b_\lo$ are $-0.969$ and $-0.614$, respectively. The differences between the one-way and two-way clustered standard errors are now smaller, reflecting a larger idiosyncratic variation across countries. Nevertheless, the one-way clustered standard errors continue to understate the underlying estimation uncertainty. Both 90\% and 68\% bootstrap confidence intervals indicate statistical significance for the estimates of both $\beta _{\mathcal{L}}$ and $b_{\lo}$. In our preferred IFE specification, the  effect of the low-frequency component  is estimated to be $-1.282$  with a 90\% bootstrap confidence interval $[-1.806, -0.835]$.   In the absence of further mitigation  policies, an additional 1°C rise in $\hat \lo_X$ would  result in a reduction of GDP growth in excess of 1.28 percentage points over the period of this low-frequency temperature increase.

Taken together, the results from the European and international panels provide convincing evidence on the differential effects of the low- and high-frequency components of temperature on real GDP growth, reaffirming the importance of accurately characterizing low-frequency temperature movements in assessing their effects on economic activity. 

\subsection{Dynamic Responses to Low Frequency Shocks}

The panel regression estimates in the preceding sections quantify the contemporaneous effect of the low- and high-frequency components of temperature on economic growth. It is often desirable to characterize the shape and the magnitude of the dynamic response of growth to a shock in the two temperature components. The long-horizon response to the low-frequency shock and its cumulative effect on output growth are also of interest.

\begin{figure*}[t!]
\caption{IRF of GDP growth to a low-frequency temperature shock}
\includegraphics[width=17cm,height=8.5cm]{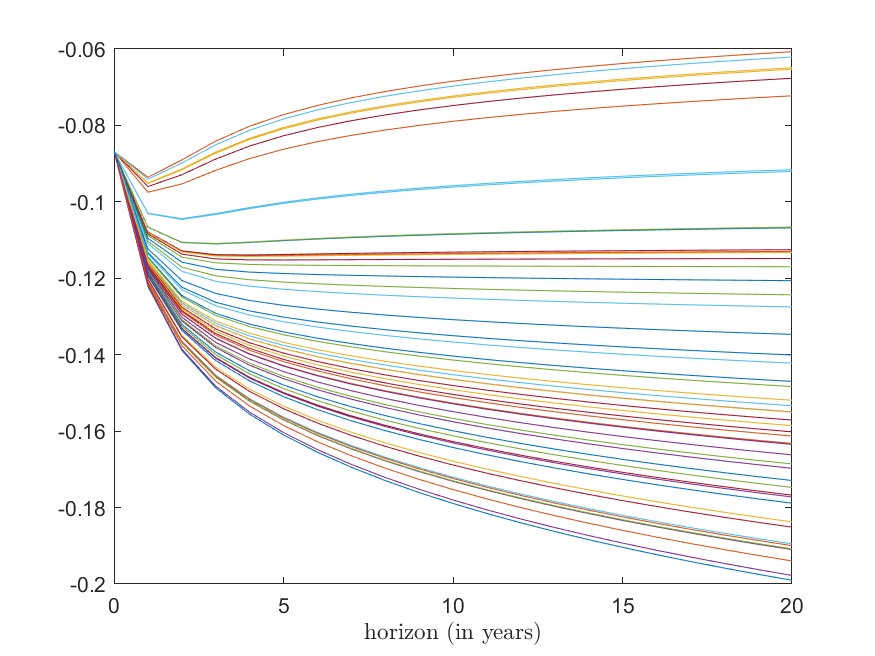}
\label{fig:IRF}

\footnotesize{ This figure plots the IRF of GDP growth to a one-standard-deviation shock to the low-frequency temperature component by country (international panel).}
\end{figure*}

In addition to the IFE model for $\Delta Y_{i}$, we also need a model for $\lo_{Xi}$ to compute the impulse response function (IRF) of $\Delta Y_{i,t+h}$ to a shock in $\lo_{it}$ at different horizons. Unfortunately, the MW decomposition is not readily amenable to this type of analysis. For this reason, we will employ again the UC model of \cite{hartl-uc} from Section 2.1 to learn about the dynamic and cumulative effects of the shocks $\epsilon_{\lo,t}$ and $\epsilon_{\hi,t}$. We first construct $\hat \lo_{it}$ and $\hat \hi_{it}$ by estimating the UC model parameters for each country in the international panel. As before, for identification purposes, we set $\sigma_{\lo}=0.1$ for all countries. We then use the dynamic IFE model to trace out the dynamic effect of $\epsilon_{\lo,t}$ on $\Delta Y_{i,t+h}$ for $h=0,1,2,\dots,20$ and $i=1,\dots,50$.\footnote{The estimated parameters are $\hat b_{\mathcal{L}}=-0.8681$, $\hat \delta _{\mathcal{H}}=0.0325$, $\hat \gamma_{\mathcal{H}}=0.2609$, and $\hat \alpha = 0.2510$.} The IRF for the low-frequency component in each country in the international panel, the most diverse of the three panels considered,  is plotted in Figure \ref{fig:IRF}. 

Several interesting observations emerge from the IRF analysis. First, unlike  shocks to the high-frequency component (not shown) that dissipate fully within 3 to 5 years, shocks to the low-frequency component have persistent effects on output growth.  This persistence is driven by the long-memory parameter $d$ which varies from 0.83 for Chile to 1.16 for Iceland (see Table \ref{tbl:intl_d} in \ref{app:AppendixTF}). This suggests that even though the variations in the low-frequency component are dominated by the high-frequency variations and its initial impact is small, its effects are long-lived and even amplified in countries where the estimate of $d$ exceeds one.   The heterogeneity in the response across countries can be better illustrated by computing the 
cumulative effect over a 20-year period. A one-standard-deviation shock to the low-frequency component in temperature will result in a reduction of real GDP growth per capita of around 1.6 (Chile) to 3.6 (Iceland) percentage points, corresponding to the top and bottom lines in Fig \ref{fig:IRF}, respectively. 
%\textcolor{red}{something seems missing throughout  is that we say how much temp changed over the sample, but nothing about output.}

\subsection{Additional Evidence on Heterogeneity and Nonlinearity}

The results regarding the average relationship between temperature and economic activity in the previous sections are based on panel data regressions.
In this section, we consider time  series estimates at the aggregate level, as well as an aggregate of state level and country level  time series estimates. Lastly, the potential for nonlinear effects is also examined.

In Section 2, we find that the national estimate of $ \lo^0_X$ is similar to the aggregate of the state level estimates $\lo^0_{X,i}$. An alternative to panel data regressions, therefore, is to use national data to estimate the time series regression shown in  (\ref{eq:coint}). Because there is only one entity, there is no need to deal with heterogeneity across units. Using national data for the U.S., we obtain an estimate of $-0.637$ for $\beta_\lo$.  Given that $\hat \lo_X$ increased by 1.693°C in the estimation sample, the point estimate of the economic loss is meaningful. However,  the  estimate  has large sampling uncertainty,  with a 90\% confidence interval of $[-1.448,0.173]$. Thus, similar to panel  estimates, the long-run effect of $\hat \lo_X$  on $\Delta Y$ is not statistically different from zero in the U.S. data.\footnote{It is easy to show that the OLS estimate from a regression of $\Delta Y$ on $\hat \lo_X$ is numerically identical to the one from a regression of $\hat \lo_Y$ on $\hat \lo_X$ so that this could be interpreted as a cointegrating model. One possible way to correct for the potential endogeneity bias in this setting is to employ a dynamic OLS (DOLS) regression. Given the smoothness of $\hat \lo_X$, however, introducing additional leads and lags of $\Delta \hat \lo_X$ results in very wide confidence intervals for $\beta_{\lo}$ due to potential collinearity.}

\begin{figure*}[ht!]
\caption{Density of OLS estimates of $\beta_{\lo}$}
\label{fig:density}
\begin{subfigure}[t]{.5\textwidth}
        \caption{U.S. States}
        \centering
        \includegraphics[width=\textwidth]{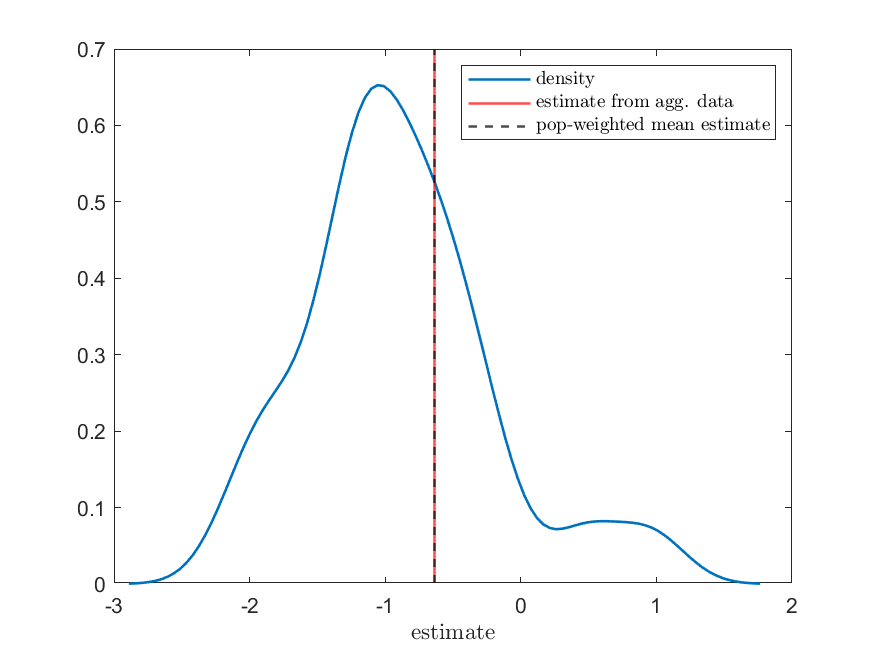}
    \end{subfigure}%
    \hfill
    \begin{subfigure}[t]{.5\textwidth}
        \caption{European Countries}
        \centering
        \includegraphics[width=\textwidth]{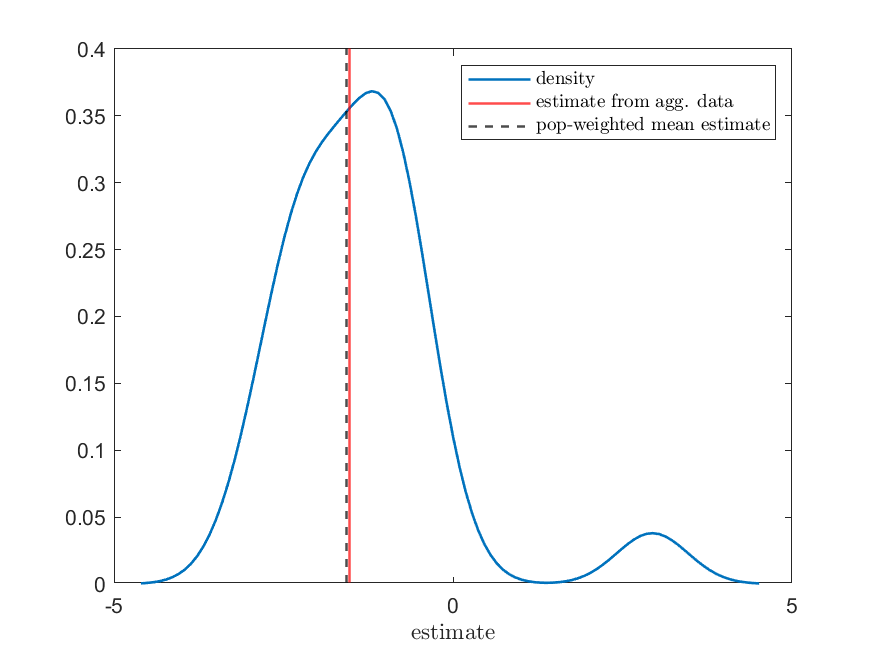}
    \end{subfigure}
\footnotesize{ This figure plots the density of $\beta_{\lo}$ from state-by-state (48 U.S. states, left chart) and country-by-country (20 European countries, right chart) time series regressions. The two charts also present the OLS estimate of $\beta_{\lo}$ from the U.S. and European aggregate regressions (solid red line), along with the population-weighted mean of the individual estimates (dashed black line).}
\end{figure*}

The pooled estimate from the panel regression (\ref{eq:static}) could mask the  heterogeneity of the economic impact  across units. Thus, we also estimate time-series regressions of $\Delta Y_{i}$ on $\mathcal{\hat{L}}_{X,i}$
and $\hat{\mathcal{H}}_{X,i}$ for each unit $i$ in the  panel under consideration.  The left chart in Figure \ref{fig:density} plots the density of the U.S. state-level  estimates for $\beta _{\mathcal{L},i}$. While the dispersion in $\beta _{\mathcal{L},i}$ estimates is large, most of the probability mass lies in the negative region.  The median of the $\beta _{\mathcal{L},i}$ estimates is $-0.954$, with the mode of the density at $-1.056$.  The population-weighted mean of $\hat\beta_\lo$ for the U.S. is $-0.638$ (dashed black line in Figure \ref{fig:density}), which is very similar to the estimate from national  data of $-0.637$ reported above (solid red line in Figure \ref{fig:density}).

 The right panel of Figure \ref{fig:density} presents a density plot of the  estimates of $\beta_{\lo,i}$ from the European sample. Compared to the U.S. (left), this density is significantly more dispersed. Except for one positive value – Ireland – all remaining point estimates are negative.\footnote{Dropping Ireland increases the pooled IFE estimate from $-0.904$ to $-0.951$.}  Ten of the negative estimates are statistically significant at 5\% nominal levels, with Newey-West $t$-statistics between $-2.085$ and $-4.364$. The mass of the density is concentrated deeper in the negative range, and the negative values are larger in absolute magnitude than those for the U.S. The mode of the density for $\{\hat\beta_{\lo,i}\}$ in Europe is $-1.195$ and  the median is $-1.359$,
 with a population-weighted average  (dashed red line) of $-1.572$. Similar to the U.S. analysis, we construct population-weighted aggregate data for these 20 European countries and estimate $\beta_\lo$ from a time-series regression using this aggregated data. The OLS estimate  of $\beta_\lo$ from this regression is $-1.528$ (solid red line in the right chart) with a 90\% confidence interval of $[-2.217,-0.839]$. This estimate is larger than the pooled estimate of $-0.904$ reported earlier, it is very close to the population weighted average, and is statistically different from zero. 

  \begin{table}
\caption{Results from Nonlinear Panel Data Regressions}
\label{tbl:nonlin}
\begin{center}

\begin{tabular}{l|ccc}
& $\beta _{\lo}$ & $\beta _{\hi}$ & $\beta _{\hi \lo} $ \\ \hline\hline
U.S. &  &  &  \\ 
estimate & $-0.338$ & $0.040$ & $-0.351$ \\ 
90\% CI & $[-0.960,0.235]$ & $[-0.400,0.465]$ & $[-0.768,0.064]$ \\ 
68\% CI & $[-0.705,0.028]$ & $[-0.213,0.282]$ & $[-0.600,-0.078]$ \\ 
\hline\hline
European &  &  &  \\ 
estimate & $-0.886$ & $0.551$ & $-0.759$ \\ 
90\% CI & $[-1.463,-0.304]$ & $[0.147,0.948]$ & $[-1.279,-0.178]$ \\ 
68\% CI & $[-1.249,-0.566]$ & $[0.280,0.764]$ & $[-1.072,-0.415]$ \\ 
\hline\hline
International &  &  &  \\ 
estimate & $-1.288$ & $0.400$ & $-0.546$ \\ 
90\% CI & $[-1.802,-0.831]$ & $[-0.020,0.780]$ & $[-1.042,-0.080]$ \\ 
68\% CI & $[-1.588,-1.006]$ & $[0.154,0.644]$ & $[-0.839,-0.246]$ \\ \hline
\end{tabular}

\end{center}

\footnotesize{The table
reports panel regression results for the static interactive fixed effects model with explanatory variables $\mathcal{\hat{H}}_{it}$, $\mathcal{\hat{L}}_{it}$, and $\mathcal{\hat{H}}_{it} \times \mathcal{\hat{L}}_{it}$. The table presents point estimates, and 90\% and 68\% bootstrap confidence intervals for the U.S. (top), European (middle) and International (bottom) panels.} \medskip

\end{table}

As surveyed in \citet{hsiang:16}, many studies have found non-linear relationships between temperature  and economic outcomes when (time-invariant) summary statistics of climate across space are interacted with temperature. \citet{kalkuhl-wenz:20} interact $X_t$ with $\Delta X_t$.  The form of nonlinearity that we introduce here differs from the forms of nonlinearity
studied in the literature (e.g., \citet{newell-prest-sexton:21}; \citet{bilal-kanzig:24}). In particular, we  include  $\hat \hi _{X,it}\times \hat \lo_{X,it}$ as an additional (time-varying) regressor in  the static specification of our preferred interactive fixed effects model. This allows for the possibility of  a new form of  nonlinearity via an interaction term of the low- and high-frequency components  with a corresponding coefficient $\beta_{\hi \lo}$.  The marginal effect of $\hat \hi_X$ on $\Delta Y$ is computed as $\hat\beta_{\hi} + \hat\beta_{\hi \lo} \bar \lo_{X,t}$, where $\bar \lo_{X,t}$ is the cross-sectional average of $\hat \lo_{X,it}$. The marginal effect of $\hat \lo_X$ on $\Delta Y$ is similarly computed as $\hat\beta_{\lo} + \hat\beta_{\hi \lo} \bar \hi_{X,t}$.

 Table \ref{tbl:nonlin} presents estimates and bootstrap confidence intervals from this nonlinear specification for the U.S. (top), European (middle) and international (bottom) panels. Compared to the results in Tables \ref{tbl:USpanel}, \ref{tbl:EUpanel} and \ref{tbl:Ipanel}, the estimates of $\beta_{\lo}$ are largely unchanged and the interaction term does not provide evidence for a nonlinear effect in the low-frequency component.  However, in our model,  nonlinearity is captured implicitly by the shape of $\hat \lo_X$ which has an upward trend since 1980.  
We find that though the estimate of $\beta_{\hi \lo}$ is also negative, the marginal effect of $\hat \lo_X$ is actually dampened when  evaluated at the cross-sectional average of $\hat \hi_X$, as this average is close to zero. 

\begin{figure*}[ht!]
\caption{Estimated nonlinear effect of $\mathcal{\hat{H}_X}$ on $\Delta Y$}
\label{fig:nonlinear}
\begin{subfigure}[t]{.5\textwidth}
        \caption{European Panel}
        \centering
        \includegraphics[width=\textwidth]{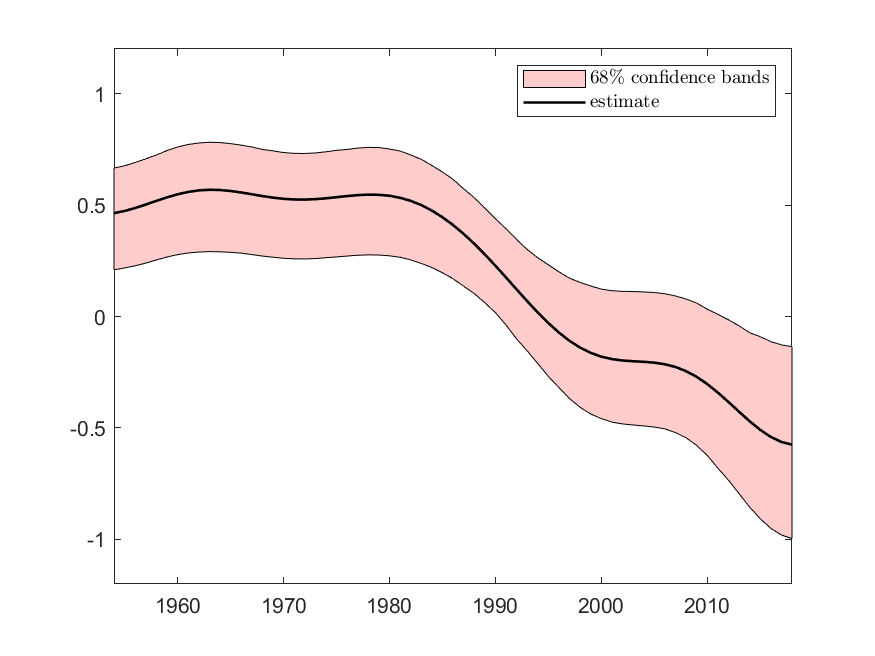}
    \end{subfigure}%
    \hfill
    \begin{subfigure}[t]{.5\textwidth}
        \caption{International Panel}
        \centering
        \includegraphics[width=\textwidth]{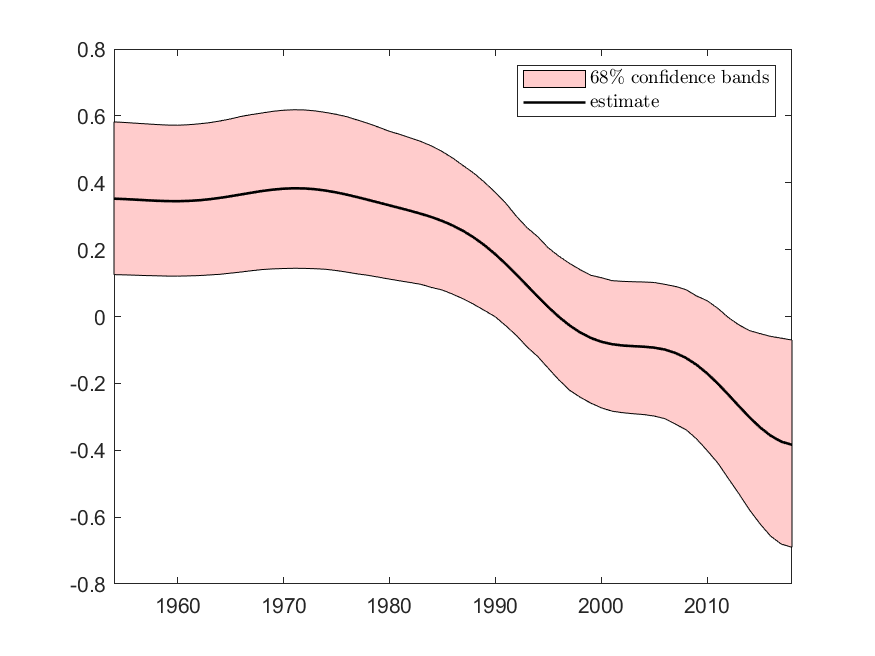}
    \end{subfigure}
\footnotesize{This figure plots the marginal effect of $\mathcal{\hat{H}_X}$ on $\Delta Y$ from the model reported in Table \ref{tbl:nonlin}, evaluated at the cross-sectional average of $ \mathcal{\hat L}_X$. The estimated marginal effect is in solid black line with 68\% bootstrap confidence bands (shaded area).}
\end{figure*}

 By contrast, the effect of the high-frequency component $\hat \hi_X$ in the European panel – which was close to zero and statistically insignificant in the linear specification – now exhibits pronounced nonlinearity, as $\hat\beta_{\hi}$ and $\hat\beta_{\hi \lo}$ appear to be significant with opposite signs. The results for $\hat \hi_X$ in the international panel also suggest some nonlinearity, while the evidence for the U.S. data is weaker. Figure \ref{fig:nonlinear} plots the marginal effect of $\hat \hi_X$ on $\Delta Y$ defined above for the European and international panels, along with 68\% bootstrap confidence bands. While the uncertainty associated with these estimates is large, there is some  evidence that the effect of $\hat\hi_X$ has switched from positive to negative since the 1980s. However,  evaluated at any $t$, the  total (linear and nonlinear) effect from a change in  $\hat \hi_X$ in both panels is noticeably smaller than the estimate for  $\beta_\lo$, which supports our premise  that low- and high-frequency changes in temperature have differential economic impacts.

\section{Concluding Remarks}
\label{sec:Sect5}

Since the late 1970s, global temperatures have registered a persistent increase that manifested in a clear break from its pre-1980 mean. In fact, 2024 was the warmest year on record, with global temperatures exceeding the pre-industrial (1850–1900) average by 1.46°C. Our analysis is premised on the fact that year-over-year changes in temperature are comprised of variations  with different degrees of persistence, and each can have  a distinct impact  on economic activity. In all three panels of data studied, the estimated  low-frequency component of temperature displays a persistent increase since the 1970s, around the same time as  the well-documented slowdown in economic growth. This informal finding guided our choice of a regression model that  allows the low- and high-frequency components of temperature to have different economic impacts and heterogeneous responses to business-cycle-related common variations in estimation and inference. Panel and time series regressions find a negative  first-order effect of the low-frequency component of temperature on economic growth that is strong and statistically significant in the international and European panels, but not in the U.S. panel.  The high-frequency component is estimated to have no first-order effect in the international data, but does exhibit some non-linear, statistically significant effect through interaction with the low-frequency component. 

\newpage

\renewcommand{\thefigure}{A\arabic{figure}}
\renewcommand{\thetable}{A\arabic{table}}
\renewcommand{\thesection}{Appendix \Alph{section}}
\setcounter{section}{0}
\setcounter{figure}{0}
\setcounter{table}{0}

\section{ Data Description}
\label{app:AppendixC}

\subsection*{U.S. Panel}

Average temperature series for the U.S. panel are from \href{https://www.ncei.noaa.gov/pub/data/cirs/climdiv/}{NOAA nClimDiv}, natively at the county–month level.\footnote{The data are available at the 30 minute (grid)-year level. We merge it with 30 minute population weights in 2020 to obtain spatial averages.} These are then (1) spatially averaged to the state level using author-constructed static population weights, and (2) averaged over time to the yearly level. The temperature series are recorded in °F but we convert them to °C to facilitate comparisons with the international data. The temperature series are then expressed as deviations from their pre-1980 mean.

Static county population weights are produced by calculating the percent of the total state population represented by each county in the 2020 IPUMS NHGIS population dataset. State population data are obtained from FRED at the yearly level. These population weights are used in computing the population-weighted averages reported in the paper.

Real Gross State Product (GSP) series are obtained from multiple sources. For data after (and including) 1997, our source is the Bureau of Economic Analysis (BEA). As they have deprecated all data before this period (both for nominal and real series), we rely on the data used in \citet{MNPRY:23} as presented in their replication package. They describe appending (1) the deflated nominal GSP by state-level CPI from 1963 to 1976, and (2) the real GSP as provided by the BEA from 1977 to 1996. In order to harmonize the \citet{MNPRY:23} series with the latest data, we normalize their price base by multiplying their data by the ratio of their 1997 value and the value for 1997 as obtained through the BEA. 

Because of missing values, Hawaii, Alaska, and the District of Columbia are dropped, giving us a cross-sectional dimension of $N=48$. The temperature data are available at the state level from 1895 to 2023 (i.e., $T=129$). After we merge temperature and real GSP growth data, the sample period is 1964--2023 and $T=60$.

\subsection*{International Panel}

Average temperature series for the international panel are from \href{https://crudata.uea.ac.uk/cru/data/hrg/cru_ts_4.07/}{CRU TS}, natively at the 30 minute grid–year level. These are then spatially averaged to the country level using author-constructed static population weights. The temperature series are in °C, and again expressed as deviations from their pre-1980 mean.

Static grid cell population weights are produced by calculating the percent of the total country population represented by each grid cell in the 2020 \href{https://sedac.ciesin.columbia.edu/data/collection/gpw-v4}{Gridded Population of the World} file. Real Gross Domestic Product (GDP) and country population is obtained from the 2023 \href{https://www.rug.nl/ggdc/historicaldevelopment/maddison/releases/maddison-project-database-2023}{Maddison Project Database}. These static (as of 2020) population weights are used in computing the population-weighted averages reported in the paper. The temperature data is available for the 1901-2022 period.

To ensure the reliability of GDP and population data, we use the \href{https://www.worldeconomics.com/DataQualityRatings/}{Data Quality Ratings} produced by World Economics. Our international panel is comprised of 50 countries with the highest combined ranking in this index. More specifically, we initially use their letter grade for GDP per capita to rank countries and then sort by the minimum of the GDP and population quality index values within these letter grades. 

The 50 countries in our international panel (with their corresponding abbreviations in parentheses) are: Switzerland (CHE), Canada (CAN), Ireland (IRL), Australia (AUS),  (GBR), Puerto Rico (PRI),\footnote{Puerto Rico is assumed to have the same data quality as the United States.} USA, Japan (JPN), New Zealand (NZL), France (FRA), Singapore (SGP), Slovenia (SVN), Taiwan (TWN), Korea (KOR), Denmark (DNK), Sweden (SWE), Netherlands (NLD), Spain (ESP), Norway (NOR), Portugal (PRT), Italy (ITA), Germany (DEU), Austria (AUT), Poland (POL), Finland (FIN), Chile (CHL), Mauritius (MUS), Hungary (HUN), Malta (MLT), Colombia (COL), North Macedonia (MKD), Mongolia (MNG), Brazil (BRA), Turkey (TUR), Malaysia (MYS), Philippines (PHL), Belgium (BEL), Luxembourg (LUX), Cyprus (CYP), Iceland (ISL), Greece (GRC), Israel (ISR), Romania (ROU), Croatia (HRV), Bulgaria (BGR), UAE, South Africa (ZAF), Uruguay (URY), Jordan (JOR), Morocco (MAR). The real GDP data for each country is at an annual frequency for the period 1952--2018. After constructing growth rates for real GDP, the resulting panel is of dimensions $T=66$ and $N=50$. 

% Finally, our European panel is a subset of the international panel and is comprised of 20 European countries with sufficient geographic variation across the European continent (no Eastern European countries are included). These countries (with their corresponding abbreviations in parentheses) are: Switzerland (CHE), Ireland (IRL), Great Britain (GBR), France (FRA), Denmark (DNK), Sweden (SWE), Netherlands (NLD), Spain (ESP), Norway (NOR), Portugal (PRT), Italy (ITA), Germany (DEU), Austria (AUT), Finland (FIN), Malta (MLT), Belgium (BEL), Luxembourg (LUX), Cyprus (CYP), Iceland (ISL), Greece (GRC). For this panel,  $T=66$ and $N=20$.

Finally, our European panel is a subset of the international panel and is comprised of 20 European countries with sufficient geographic variation across the European continent (no Eastern European countries are included). These countries (with their corresponding abbreviations in parentheses) are: Switzerland, Ireland, Great Britain, France, Denmark, Sweden, Netherlands, Spain, Norway, Portugal, Italy, Germany, Austria, Finland, Malta, Belgium, Luxembourg, Cyprus, Iceland, Greece. For this panel,  $T=66$ and $N=20$.

\subsection*{Summary Statistics}

Table \ref{tbl:summary} reports summary statistics for these three panels. For all of these datasets, the temperature has increased by around 1.2°C at the end of the sample while the GDP growth has slowed significantly over this period.  

\bigskip

\begin{table}[ht]

\begin{center}
\caption{Summary Statistics for Temperature ($X$) and Growth ($\Delta Y$) for Different Panels.}

\label{tbl:summary}

\begin{tabular}{c|cc|cc|cc|cc}
& \multicolumn{2}{|c}{U.S.} & \multicolumn{2}{|c}{Europe} & 
\multicolumn{2}{|c}{Int'l (top 50)} & \multicolumn{2}{|c}{Int'l (rest)} \\ 
\hline
& $X$ & $\Delta Y$ & $X$ & $\Delta Y$ & $X$ & $\Delta Y$ & $X$ & $\Delta Y$
\\ \hline
ave$_{i}$ave$_{t}$ & $0.524$ & $1.910$ & $0.475$ & $2.744$ & $0.470$ & $2.816
$ & $0.329$ & $1.862$ \\ 
ave$_{i}$ave$_{1:10}$ & $-0.090$ & $3.591$ & $0.117$ & $3.658$ & $0.140$ & $%
3.507$ & $0.056$ & $2.069$ \\ 
ave$_{i}$ave$_{T:T-9}$ & $1.246$ & $1.343$ & $1.243$ & $0.501$ & $1.209$ & $%
1.382$ & $0.848$ & $2.073$ \\ 
$\min_{i}$ave$_{t}$ & $0.040$ & $0.852$ & $0.195$ & $1.840$ & $0.191$ & $%
1.530$ & $-0.133$ & $-2.102$ \\ 
$\max_{i}$ave$_{t}$ & $0.955$ & $2.683$ & $0.706$ & $4.448$ & $0.839$ & $%
5.452$ & $0.725$ & $5.334$ \\ 
$\min_{i}\min_{t}$ & $-1.872$ & $-10.89$ & $-1.859$ & $-19.62$ & $-1.859$ & $%
-29.78$ & $-1.224$ & $-85.50$ \\ 
$\max_{i}\max_{t}$ & $3.205$ & $29.74$ & $2.528$ & $19.48$ & $3.548$ & $27.51
$ & $2.754$ & $83.67$ \\ 
ave$_{i}$var$_{t}$ & $0.615$ & $9.431$ & $0.579$ & $9.266$ & $0.449$ & $%
14.334$ & $0.210$ & $44.20$ \\ 
$\min_{i}$var$_{t}$ & $0.338$ & $4.274$ & $0.213$ & $3.182$ & $0.106$ & $%
2.759$ & $0.052$ & $4.126$ \\ 
$\max_{i}$var$_{t}$ & $1.266$ & $44.09$ & $1.124$ & $41.76$ & $1.124$ & $%
52.76$ & $0.627$ & $482.4$ \\ \hline
\end{tabular}

\end{center}
\footnotesize{The temperature ($X$) is measured in Celsius as deviations from the pre-1980 mean. The GDP growth rate ($\Delta Y$) is in percent. 'U.S.', 'Europe' and 'Int'l (top 50)' are the three panels described above and used in this paper. 'Int'l (rest)' are the countries from rank 51 to 133 in the \href{https://www.worldeconomics.com/DataQualityRatings/}{Data Quality Ratings}. The reported summary statistics use the following abbreviations. 'ave$_i$' and 'ave$_t$' denote averages across the cross-sectional ($i$) and time ($t$) dimensions. 'ave$_{1:10}$' and 'ave$_{T:T-9}$' denote averages for the first 10 and last 10 time series observations. '$\min_{i}$' ('$\max_{i}$') and '$\min_{t}$' ('$\max_{t}$') denote the minimum (maximum) values across $i$ and $t$, respectively. Finally, var$_{t}$ is the time-series average of the series.
}
\end{table}

One potential drawback of our analysis is the absence of low-income countries in the selected international panel. In fact, it is reasonable to conjecture that the impact from global warming is expected to be stronger for low-income countries that are located in regions with unfavorable climate. Our decision to exclude most of these countries from the international panel was driven by the lack of reliable data for economic activity for countries beyond the top 50 in the data quality ratings. To illustrate this, the last two columns of Table \ref{tbl:summary} report summary statistics for countries ranked between 51 and 133, for which data is available. Indeed, the GDP growth exhibits an extreme variability that is likely driven by other extraneous factors that are not controlled for in our regressions. Somewhat surprisingly, however, the temperature data is characterized with less variability and more limited evidence of global warming than the three panels used in our empirical analysis.

\section{Additional Empirical Evidence}
\label{app:AppendixTF}

\renewcommand{\thetable}{B\arabic{table}}
\renewcommand{\thefigure}{B\arabic{figure}}
\setcounter{table}{0}
\setcounter{figure}{0}

\begin{figure}[t]
\caption{Aggregate Variations by Components for Europe}
\label{fig:fhsw5_europe}
\begin{subfigure}[b]{0.5\textwidth}
    \caption{GDP Growth}
    \includegraphics[width=1.1\textwidth]{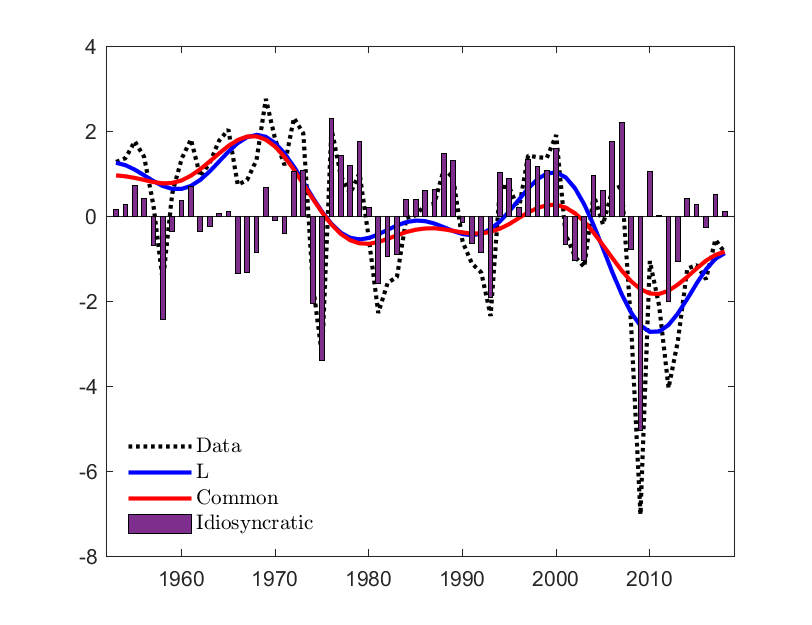}
\end{subfigure}%
~ 
\begin{subfigure}[b]{0.5\textwidth}
    \centering
    \caption{Temperature}
    \includegraphics[width=1.1\textwidth]{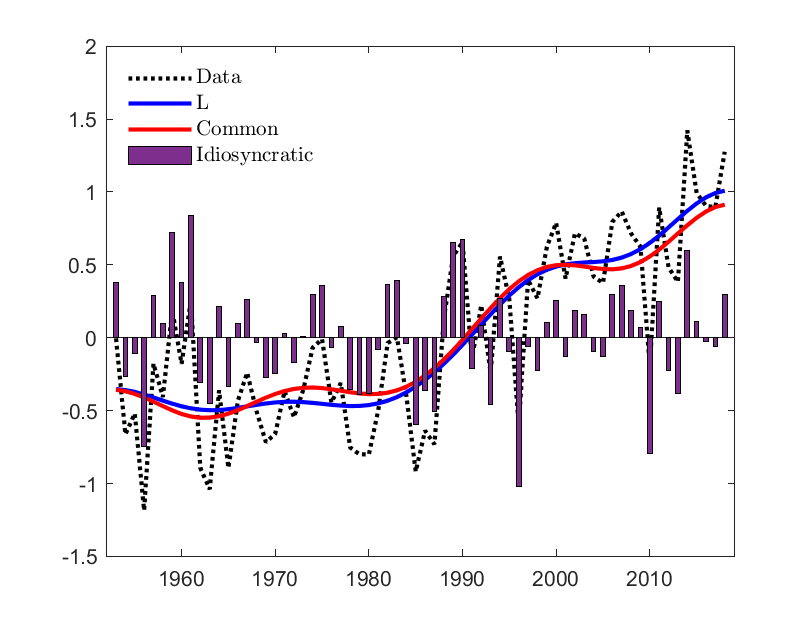}
\end{subfigure}
\vfill 
%\footnotesize{See notes to Figure \ref{fig:fhsw5_us}.}
\footnotesize{This figure presents the actual series (black dashed line), its low-frequency component estimated using MW8 (in blue),  as well as its decomposition into  a common component (in red) and idiosyncratic error (in brown bars) using the framework of \citet{fhsw-jpe:22}).  The two variables are aggregate GDP growth $\Delta Y$ (left) and aggregate temperature $X$ (right). The sample period is 1952-2018.}
\end{figure}

In this appendix, we present an additional figure and table for the international panels. Figure \ref{fig:fhsw5_europe} plots a decomposition of the variations for temperature and output growth for the 20 European countries. The sample period is 1953--2018.

% \noindent Table \ref{tbl:europe_plot_2} summarizes the European data in terms of the UC parameters estimated over the full sample, 1901-2022 with $\sigma_\lo=0.1$. The  countries whose $\hat \lo_X$  experienced the smallest change over this period are Ireland, Great Britain, Greece, and Iceland, while the countries with the largest changes are  Austria, Switzerland, Luxembourg, and France.  The innovations to the high-frequency component are largest in the Scandinavian countries and smallest in Malta, Portugal, and Spain. 

% \begin{table}[ht]
% %% plotd.tex
% \caption{UC Estimates for 20 European Countries Ordered by $\sqrt{\hat \nu}=\hat\sigma_\hi/ \sigma_\lo$}  
% \label{tbl:europe_plot_2}
% \vspace*{-.2in}
% \begin{center}
% % 1901-2022 \qquad $T=122$ \qquad $\sigma_\lo=0.2$
 
% \medskip
% \begin{tabular}{lllll|lllll}

% & country & $\hat d$ & $\Delta^T \hat \lo_{X}$ & $\hat \sigma_\hi$ 
% & country & $\hat d$ & $\Delta^T \hat \lo_{X}$ & $\hat \sigma_\hi$  \\ \hline
% & \multicolumn{4}{c}{Bottom half} & \multicolumn{4}{c}{ Top half} 
% \\ \hline

% \input tables/plot_d10_europe.tex
% \end{tabular}
% \end{center}
% {\footnotesize The sample period is 1901--2022 ($T=122$). $X$ is the deviation of temperature (in Celsius) from the pre-1980 mean. $\Delta^T \hat \lo_X=\hat \lo_{X,2022}-\hat \lo_{X,1901}$, $\lo_X$ estimated by the UC model with $\sigma_\lo=0.1$, $\hat d$ is the long memory parameter, $\hat\sigma_\hi$ is the innovation to the high-frequency component. See \ref{app:AppendixC} for the country name abbreviations.}

% \end{table}

Table \ref{tbl:intl_d} reports the UC model estimates and $\Delta^T \hat \lo_X=\hat \lo_{X,2018}-\hat \lo_{X,1953}$ for the 50 countries in the international panel. The sample period is 1953--2018. The  countries whose $\hat \lo_X$  experienced the largest change over this period appear to be European countries.  The innovations to the high-frequency component are largest in the Scandinavian countries and smallest in the southern hemisphere, including southern European countries.

\begin{table}
\caption{UC Estimates for 50 Countries (International Panel) Ordered by $\sqrt{\hat \nu}=\hat\sigma_\hi/ \sigma_\lo$}  
\label{tbl:intl_d}
\vspace*{-.2in}
\begin{center}
% 1901-2022 \qquad $T=122$ \qquad $\sigma_\lo=0.2$
 
\medskip
\begin{tabular}{lllll|lllll}

& country & $\hat d$ & $\Delta^T \hat \lo_{X}$ & $\hat \sigma_\hi$ 
& country & $\hat d$ & $\Delta^T \hat \lo_{X}$ & $\hat \sigma_\hi$  \\ \hline
& \multicolumn{4}{c}{Bottom half} & \multicolumn{4}{c}{ Top half} 
\\ \hline

&   PHL & 0.844 & 0.769 &0.174&   ISR & 1.148 & 1.531 &0.439&\\
 &   SGP & 0.977 & 1.032 &0.175&   JOR & 1.154 & 1.538 &0.447&\\
 &   MYS & 0.935 & 0.867 &0.175&   KOR & 1.027 & 1.260 &0.450&\\
 &   PRI & 1.003 & 1.261 &0.212&   GBR & 0.994 & 0.977 &0.453&\\
 &   MUS & 1.020 & 1.201 &0.215&   CYP & 1.144 & 1.438 &0.463&\\
 &   BRA & 0.936 & 1.120 &0.225&   FRA & 1.094 & 1.548 &0.476&\\
 &   AUS & 0.993 & 1.180 &0.226&   ISL & 1.156 & 0.963 &0.492&\\
 &   TWN & 0.977 & 1.153 &0.258&   HRV & 1.144 & 1.915 &0.493&\\
 &   CHL & 0.827 & 0.594 &0.268&   SVN & 1.142 & 1.914 &0.494&\\
 &   MLT & 1.089 & 1.591 &0.280&   CHE & 1.124 & 1.813 &0.495&\\
 &   COL & 0.846 & 0.593 &0.287&   ROU & 1.135 & 1.603 &0.514&\\
 &   ZAF & 1.011 & 1.339 &0.296&   TUR & 1.103 & 1.283 &0.522&\\
 &   URY & 0.833 & 0.786 &0.311&   CAN & 0.992 & 1.104 &0.536&\\
 &   ITA & 1.083 & 1.640 &0.314&   AUT & 1.132 & 1.930 &0.538&\\
 &   PRT & 1.058 & 1.288 &0.323&   HUN & 1.141 & 1.800 &0.567&\\
 &   ESP & 1.087 & 1.547 &0.328&   LUX & 1.121 & 1.641 &0.582&\\
 &   ARE & 1.067 & 1.316 &0.336&   BEL & 1.109 & 1.470 &0.589&\\
 &   USA & 0.997 & 1.068 &0.342&   NLD & 1.092 & 1.372 &0.606&\\
 &   MAR & 1.042 & 1.110 &0.355&   DEU & 1.107 & 1.599 &0.607&\\
 &   GRC & 1.083 & 1.075 &0.359&   MNG & 1.122 & 1.759 &0.650&\\
 &   MKD & 1.098 & 1.466 &0.361&   POL & 1.098 & 1.490 &0.666&\\
 &   NZL & 0.872 & 0.695 &0.373&   DNK & 1.080 & 1.335 &0.700&\\
 &   IRL & 0.855 & 0.478 &0.374&   NOR & 1.054 & 1.112 &0.722&\\
 &   JPN & 1.070 & 1.396 &0.384&   SWE & 1.077 & 1.225 &0.761&\\
 &   BGR & 1.118 & 1.554 &0.431&   FIN & 1.114 & 1.451 &0.879&\\
 \hline

\end{tabular}
\end{center}
{\footnotesize The sample period is 1953--2018 ($T=66$). $X$ is the deviation of temperature (in Celsius) from the pre-1980 mean. $\Delta^T \hat \lo_X=\hat \lo_{X,2018}-\hat \lo_{X,1953}$, $\lo_X$ estimated by the UC model with $\sigma_\lo=0.1$, $\hat d$ is the long memory parameter, $\hat\sigma_\hi$ is the innovation to the high-frequency component. See \ref{app:AppendixC} for the country name abbreviations.}

\end{table}

\section{Alternative Estimators of  $\lo^0$}
\label{app:AppendixL}

\renewcommand{\thetable}{C\arabic{table}}
\renewcommand{\thefigure}{C\arabic{figure}}
\setcounter{table}{0}
\setcounter{figure}{0}

The MWq procedure of \citet{mueller-watson-ecma:08,mueller-watson:17,mueller-watson:22} is described in the main text (Section 2.2). The HP($\lambda$) filter of \citet{hodrick-prescott}  uses a two-sided smoother with parameter $\lambda$ to approximate the slow-moving component.
\citet{phillips-jin:21} and \citet{phillips-shi:21}  show that boosting, or repeated application of  the HP-filter can  improve trend estimates if the data are integrated of order $d \in [0,1]$.
Specifically,  `bHP' produces   $\lo=Z- \hi$, where $ \hi=(I_n-S(\lambda)^m Z$, $S(\lambda)=(I_T+\lambda D_2D_2')^{-1}$, $D_2'$ is a $(T-2)\times T$ submatrix of the $T\times T$ Toeplitz matrix whose first row is $(1,-2,1,0_{T-3})$.
 Without iteration, bHP($\lambda$) is the HP($\lambda$), but the bHP endogenously adjusts $\lambda$ at each step of the iteration. Building on this work,
 \citet{phillips-longmemory} establish the conditions under which bHP can also be applied to data with long-range dependence.

\citet{hamilton:18} forcefully argues that the HP filters create spurious dynamics. The proposed JH(h,p) procedure constructs $\hi_t=Z_{t+h}-\text{proj}(Z_{t+h}|Z_t,Z_{t-1},\ldots,Z_{t-p}),$   which is the unpredicted value of $Z_{t+h}$  based on information at time $t$ and earlier. This amounts to taking $\hat\lo$ to be  the fit from  regressing  $Z_{t+h}$ on  values of $Z_t$ and its $p$ lags.  The $\hat \lo$ produced by JH is  a filtered estimate of $\lo^0$ which will be more variable than   MWq and bHP because these two are based on two-sided smoothing,  and thus   subject to `look-ahead' bias.

\subsection{A Monte Carlo Exercise for Estimation of $\lo^0_X$}
The procedures described in Section \ref{sec:Sect2} are designed for trend filtering of economic data. We do not know how accurate they are for estimating temperature trends, and for this, we need to  know the true underlying low-frequency component which is never observed. We thus design  a  Monte-Carlo exercise calibrated to several temperature series to evaluate the total (approximation plus sampling) mean-squared error of the procedures.

We use the UC model to generate data in the Monte-Carlo exercise, and the  parameters  for that model are  $(d,\sigma_\lo,\sigma_\hi,a_1)$. We select the following seven states which have  different configurations of $(\hat d,\hat\sigma_\hi$):  California (CA), Florida (FL), Illinois (IL), Massachusetts (MA), North Dakota (ND), New York (NY), and Washington (WA). 
As seen from Table \ref{tbl:sbs_plot_d},  ND has the largest $\hat\sigma_H$ while CA and FL have some of the smallest. WA and MA have similar $\hat\sigma_\hi$ but different values of $\hat d$.
    Except for IL and ND, which are located in the interior of the country, the other four  states are along the coast.  In each of the $S$ simulations of data for  state $i$,   $ \lo^0_{it}$ is fixed to be $\hat \lo_{it}$, the state-level estimate  with $\sigma_{\lo}=0.1$ or 0.2.  The state-specific  parameters $(\hat a_{i1},\hat\sigma_{i,\hi})$ are then used to generate  $\hi^s_{it}$, after which  $X_{it}^s=\hat \lo_{it}+\hat \hi_{it}^s$ is constructed.  The simulated series $\{X_{it}^s\}$  is used by different procedures to estimate  $\{\lo_{it}^s\}$. 

 Table \ref{tbl:montecarlo} reports the in-sample root mean-squared error (RMSE).  The top panel gives results for  $\sigma_\lo=0.1$, which will simulate a smoother  $\lo^0$  compared to $\sigma_\lo=0.2$, which is shown in the bottom panel. Values for $\hat d$ and $\hat\sigma_\hi$ used in the calibration are given in the first row.  In both cases, the RMSE  is largest for ND which is the state  with the largest  $\hat\sigma_\hi$, and  smallest for  FL which has the smallest $\hat\sigma_\hi$. Indeed, comparing results across states, the errors are increasing in $\hat\sigma_\hi$. In the top panel where $\sigma_{\lo}=0.1$,   MW(8) and HP(100) have the smallest RMSE,     while JH(h,p) has the largest errors.   
 In the bottom panel where $\sigma_\lo=0.2$,  MW (either MW(8) or MW(12)) and HP(100) still  have the smallest RMSE. Though HP(100) performs well, it is unclear how to justify the choice of $\lambda=100$ because, as noted above, the estimates  of $d$ are always smaller than the value of two that is implicit in the HP filter.

\begin{table}[ht]

\begin{center}
\caption{Monte Carlo: In-Sample MSE for $X_i$}

\label{tbl:montecarlo}

\begin{tabular}{l|c|c|c|c|c|c|c}
& 
\multicolumn{1}{c|}{CA}&
\multicolumn{1}{c|}{FL} &
\multicolumn{1}{c|}{IL} &
\multicolumn{1}{c|}{MA} &
\multicolumn{1}{c|}{ND} &
\multicolumn{1}{c|}{NY} &
\multicolumn{1}{c}{WA} \\

\hline
\multicolumn{7}{c}{DGP: $\sigma_\lo=0.1$ } \\ \hline

 &(1.02,0.41) &(1.04,0.40) &(0.91,0.76) &(1.09,0.56) &(0.93,0.98) &(1.05,0.59) &(0.87,0.55)\\ \hline
          MW(8) & 1.486 &  1.442 &  2.541 &  1.909 &  3.662 &  1.918 & 2.010 \\
         MW(12) & 1.663 &  1.526 &  3.047 &  2.156 &  4.384 &  2.244 & 2.369 \\
         MW(16) & 1.851 &  1.710 &  3.477 &  2.396 &  5.004 &  2.514 & 2.693 \\
        JH(4,4) & 2.619 &  2.517 &  3.089 &  3.303 &  4.104 &  3.350 & 2.589 \\
        JH(6,6) & 2.883 &  2.822 &  3.109 &  3.504 &  4.156 &  3.471 & 2.553 \\
        JH(8,8) & 3.340 &  3.156 &  3.233 &  3.995 &  4.302 &  3.904 & 2.655 \\
      bHP(6.25) & 2.409 &  2.205 &  4.564 &  3.148 &  6.486 &  3.312 & 3.496 \\
        HP(100) & 1.574 &  1.455 &  2.903 &  2.022 &  4.177 &  2.114 & 2.255 \\

\hline
\multicolumn{7}{c}{DGP: $\sigma_\lo=0.2$ } \\ \hline

 &(0.80,0.38) &(0.80,0.36) &(0.69,0.75) &(0.87,0.53) &(0.74,0.97) &(0.83,0.56) &(0.66,0.53)\\ \hline
          MW(8) & 1.823 &  1.821 &  2.549 &  2.125 &  3.618 &  2.061 & 2.092 \\
         MW(12) & 1.818 &  1.697 &  3.024 &  2.208 &  4.308 &  2.292 & 2.324 \\
         MW(16) & 1.848 &  1.792 &  3.423 &  2.292 &  4.912 &  2.433 & 2.585 \\
        JH(4,4) & 2.548 &  2.511 &  3.116 &  3.230 &  4.198 &  3.288 & 2.640 \\
        JH(6,6) & 2.985 &  2.946 &  3.153 &  3.532 &  4.239 &  3.493 & 2.613 \\
        JH(8,8) & 3.453 &  3.226 &  3.278 &  4.097 &  4.391 &  3.982 & 2.740 \\
      bHP(6.25) & 2.092 &  1.933 &  4.454 &  2.882 &  6.355 &  3.099 & 3.273 \\
        HP(100) & 1.644 &  1.594 &  2.871 &  2.000 &  4.104 &  2.099 & 2.196 \\

\hline

\end{tabular}

\end{center}
\footnotesize{Reported are results based on 5,000 replications. The seven states are California (CA), Florida (FL), Illinois (IL), Massachusetts (MA), North Dakota (ND), New York (NY), and Washington (WA). For each state $i=1,\ldots,7$ and for each replication $s$, data $\hat X^s_i=\hat \lo_{Xi,i}+\hat  \hi_{X,i}^s$ are obtained by generating $\hat \hi_{X,i}$ using  parameter estimates $(\hat d, \hat a_1,\hat\sigma^2_\hi)$ for state $i$  with $\sigma_\lo$ held fixed. Results are reported for $\sigma_\lo=0.1$ (top panel)  and 0.2 (bottom panel).
}
\end{table}

%\newpage
\section{Simulation Evidence for Panel Data Regressions}
\label{app:AppendixB}
\renewcommand{\thetable}{D\arabic{table}}
\setcounter{table}{0}

This section provides simulation evidence for assessing the properties of
the estimation and inference methods used in Section \ref{sec:Sect4} of the paper. We
consider a dynamic panel data model with (i) interactive fixed effects (IFE) and
(ii) individual fixed effects (FE). As in the paper, we are interested in the
static and dynamic effects of the low-frequency ($\mathcal{L}$) and
high-frequency ($\mathcal{H}$) components.

The true processes $\mathcal{L}_{it}$ and $\mathcal{H}_{it}$ are assumed to
follow the decomposition described in Section 2.1 for each $i=1,...,N$. The
parameters that describe the dynamics of these processes are estimated from
the U.S. state temperature data using the \citet{hartl-uc} estimator for $\sigma_\lo=0.2,0.1$, and $0.02$. Given the strong
factor structure that we document in the main text, we postulate that both $%
\mathcal{L}_{it}$ and $\mathcal{H}_{it}$ are driven by one common factor. We
keep the common factor in the state low-frequency components fixed (at their
estimated value) in the simulation with variation in the simulated $%
\mathcal{L}_{it}$ coming only from drawing idiosyncratic errors from a
multivariate Gaussian distribution with mean zero and a covariance matrix $%
\Sigma _{\mathcal{L}}$, calibrated to the data. For the high-frequency
common component, we further decompose it into a deterministic component --
estimated by MWq with $q=28$, which is intended to approximate the ENSO cycle -- and an AR(1) cyclical component. We keep the deterministic
part fixed in simulations and generate AR(1) processes from a Gaussian
distribution with AR and variance parameters calibrated to the data. We
then build the common factor structure in $\mathcal{H}_{it}$ using the
estimated loadings from the data and generate idiosyncratic errors from a
multivariate Gaussian distribution with mean zero and variance $\Sigma _{%
\mathcal{H}}$. All these steps produce simulated data $\mathcal{\tilde{L}}%
_{it}$ and $\mathcal{\tilde{H}}_{it}$ as well as $\tilde{X}_{it}=\mathcal{%
\tilde{L}}_{it}+\mathcal{\tilde{H}}_{it}$.

For the FE setup, we generate data for $Y_{it}$ from the
model%
\begin{equation*}
\Delta \tilde{Y}_{it}=\hat{\alpha}\Delta \tilde{Y}_{i,t-1}+\hat{b}_{\mathcal{L}}%
\mathcal{\tilde{L}}_{it}+\hat \delta_{\hi} \Delta \mathcal{\tilde{H}}_{it} + \hat{\gamma}_{\mathcal{H}}\mathcal{\tilde{H}}_{i,t-1}+%
\tilde{u}_{Y,it},
\end{equation*}%
where $\tilde{u}_{Y,it}\sim N(0,\hat{\Sigma}_{u})$, and $\hat{\alpha}$, $%
\hat{b}_{\mathcal{L}}$, $\hat \delta_{\hi}$, $\hat{\gamma}_{\mathcal{H}}$ and $\hat{\Sigma}_{u}$ are the sample estimates from a dynamic panel data model with individual
fixed effects. For the IFE specification, the data for 
$Y_{it}$ is instead generated from the model
\begin{equation*}
\Delta \tilde{Y}_{it}=\hat{\alpha}\Delta \tilde{Y}_{i,t-1}+\hat{b}_{\mathcal{L}}%
\mathcal{\tilde{L}}_{it}+\hat \delta_{\hi} \Delta \mathcal{\tilde{H}}_{it} + \hat{\gamma}_{\mathcal{H}}\mathcal{\tilde{H}}_{i,t-1}+%
\hat{\lambda}_{i}\hat{F}_{t}+\tilde{u}_{Y,it},
\end{equation*}%
where $\tilde{u}_{Y,it}\sim N(0,\hat{\Sigma}_{u})$, and $\hat{\alpha}$, $%
\hat{b}_{\mathcal{L}}$, $\hat \delta_{\hi}$, $\hat{\gamma}_{\mathcal{H}}$, $\hat{\lambda}_{i}$%
, $\hat{F}_{t}$ and $\hat{\Sigma}_{u}$ are now the sample estimates from a
dynamic panel data model with interactive fixed effects as in Bai (2009). 

In order to preserve the main features of the data, we keep $\hat{\lambda}%
_{i}\hat{F}_{t}$ fixed in the simulations. We also remain agnostic about
the fact that $\mathcal{\tilde{L}}_{it}$ and $\mathcal{\tilde{H}}_{it}$ are
generated by the Hartl procedure and, instead, we estimate them using the
\cite{mueller-watson-ecma:08,mueller-watson:17,mueller-watson:22} method, MWq, for several values of the tuning parameter $q$. This allows us to assess the effects of uncertainty about $\mathcal{L}$ and $\mathcal{H}$ on the estimation of the
parameters of interest. All of the simulated data is of dimension $(T=60,N=48)$. The
results are based on 1,000 Monte Carlo replications.

In addition to estimation, we also evaluate the coverage properties of
several inference procedures. Since $N$ is relatively small, it would be
desirable for the inference procedure to account for the uncertainty in
estimating $F_{it}$ and $\lambda _{it}$. For this reason, and given the
complex dependence structure in the IFE model, we use
a fixed-design bootstrap where the only source of variation comes from
resampling the errors $\tilde{u}_{Y,it}$. We use $399$ bootstrap samples
to re-estimate all of the model parameters and construct 90\% confidence
intervals using the percentile method. We use this bootstrap
method in both the FE and the IFE designs.

In the individual fixed effects model, standard errors are routinely produced
by clustering at state level. But given the strong factor structure
documented in Section \ref{sec:Sect4}, valid inference necessitates two-way clustering
in order to accommodate the more complex dependence structure in these
models. While the IFE model explicitly takes this
factor structure into account, the FE model leaves it in
the errors and the inference procedure needs to be adjusted to reflect this
additional source of heterogeneity. For these reasons, we include in the
simulations coverage of 90\% confidence intervals (based on standard normal asymptotic approximation) using one-way (within $i$) cluster-robust standard errors (referred to
as \textquotedblleft asy1\textquotedblright in the Table \ref{tab:tabB2}) and two-way
cluster-robust standard errors as in \citet{CGM2011}
(referred to as \textquotedblleft asy2\textquotedblright\ in the Table \ref{tab:tabB2}).

The results can be summarized as follows. For the IFE
model (Table \ref{tab:tabB1}), the estimates exhibit little bias. The coverage
rates of the 90\% bootstrap confidence intervals are close to the nominal
level although they undercover for specifications where the variation in $%
X_{it}$ is dominated by the high-frequency component and $q$ is relatively
large. For the FE model (Table \ref{tab:tabB2}), we observe a
 bigger (downward) bias of the estimates although these are
relatively minor given the small $N$ and $T$ dimensions. The most striking result is
the severe under-coverage of confidence intervals based on one-way clustered
standard errors. The two-way clustering substantially improves the coverage
of the 90\% asymptotic confidence intervals with the bootstrap method delivering
further improvements. In summary, this simulation evidence is supportive and
reassuring for the results presented in our empirical analysis.\pagebreak

\begin{table}
\caption{
Results for the IFE model}
\label{tab:tabB1}

\begin{center}
\begin{tabular}{c|c|ccc|ccc|ccc}
&  & \multicolumn{3}{|c|}{$q=4$} & \multicolumn{3}{|c|}{$q=8$} & 
\multicolumn{3}{|c}{$q=12$} \\ \hline
coeff & true & estim & sd & boot & estim & sd & boot & estim & sd & boot \\ 
\hline
$\sigma _{\mathcal{L}}=0.2$ &  &  &  &  &  &  &  &  &  &  \\ \hline
$b_{\mathcal{L}}$ & $-0.281$ & $-0.265$ & $0.318$ & $0.890$ & $-0.242$ & $%
0.291$ & $0.866$ & $-0.220$ & $0.267$ & $0.864$ \\ 
$\delta _{\mathcal{H}}$ & $-0.228$ & $-0.231$ & $0.170$ & $0.898$ & $-0.229$
& $0.178$ & $0.892$ & $-0.232$ & $0.188$ & $0.882$ \\ 
$b_{\mathcal{H}}$ & $-0.119$ & $-0.124$ & $0.249$ & $0.885$ & $-0.120$ & $%
0.272$ & $0.888$ & $-0.127$ & $0.292$ & $0.887$ \\ 
$\alpha $ & $0.164$ & $0.141$ & $0.040$ & $0.865$ & $0.142$ & $0.040$ & $%
0.862$ & $0.142$ & $0.040$ & $0.865$ \\ \hline
$\sigma _{\mathcal{L}}=0.1$ &  &  &  &  &  &  &  &  &  &  \\ \hline
$b_{\mathcal{L}}$ & $-0.087$ & $-0.119$ & $0.376$ & $0.857$ & $-0.134$ & $%
0.332$ & $0.856$ & $-0.148$ & $0.300$ & $0.852$ \\ 
$\delta _{\mathcal{H}}$ & $-0.267$ & $-0.254$ & $0.177$ & $0.882$ & $-0.255$
& $0.186$ & $0.875$ & $-0.254$ & $0.196$ & $0.886$ \\ 
$b_{\mathcal{H}}$ & $-0.198$ & $-0.196$ & $0.265$ & $0.868$ & $-0.198$ & $%
0.286$ & $0.870$ & $-0.194$ & $0.307$ & $0.880$ \\ 
$\alpha $ & $0.164$ & $0.142$ & $0.040$ & $0.874$ & $0.142$ & $0.040$ & $%
0.878$ & $0.142$ & $0.041$ & $0.881$ \\ \hline
$\sigma _{\mathcal{L}}=0.02$ &  &  &  &  &  &  &  &  &  &  \\ \hline
$b_{\mathcal{L}}$ & $-0.045$ & $-0.084$ & $0.384$ & $0.867$ & $-0.106$ & $%
0.329$ & $0.870$ & $-0.128$ & $0.298$ & $0.856$ \\ 
$\delta _{\mathcal{H}}$ & $-0.266$ & $-0.257$ & $0.169$ & $0.893$ & $-0.257$
& $0.177$ & $0.894$ & $-0.256$ & $0.184$ & $0.893$ \\ 
$b_{\mathcal{H}}$ & $-0.197$ & $-0.197$ & $0.247$ & $0.890$ & $-0.197$ & $%
0.266$ & $0.898$ & $-0.196$ & $0.287$ & $0.892$ \\ 
$\alpha $ & $0.164$ & $0.142$ & $0.041$ & $0.874$ & $0.142$ & $0.041$ & $%
0.875$ & $0.142$ & $0.041$ & $0.877$ \\ \hline
\end{tabular}

\end{center}

\footnotesize{ The table
reports the true value of the parameters (\textquotedblleft true\textquotedblright), the corresponding average
estimates (\textquotedblleft estim\textquotedblright) over 1,000 Monte Carlo simulations, their standard deviations (\textquotedblleft sd\textquotedblright) and
the coverage rates of 90\% bootstrap confidence intervals (\textquotedblleft boot\textquotedblright). The results are
reported for 3 values of $\sigma _{\mathcal{L}}$ (the standard deviation of
the low-frequency component $\mathcal{L}$) and 3 values of $q$ for the MWq
procedure, which is used to estimate the latent $\mathcal{L}$.}
\end{table}
\pagebreak

\begin{table}[ht]
\caption{
Results for the FE model}
\label{tab:tabB2}
\begin{center}
\begin{tabular}{c|c|ccccc|ccccc}
&  & \multicolumn{5}{|c|}{$q=4$} & \multicolumn{5}{|c}{$q=8$} \\ \hline
coeff & true & estim & sd & asy1 & asy2 & boot & estim & sd & asy1 & asy2 & 
boot \\ \hline
$\sigma _{\mathcal{L}}=0.2$ &  &  &  &  &  &  &  &  &  &  &  \\ \hline
$b_{\mathcal{L}}$ & $-0.948$ & $-0.833$ & $0.562$ & $0.217$ & $0.836$ & $%
0.856$ & $-0.765$ & $0.531$ & $0.198$ & $0.819$ & $0.832$ \\ 
$\delta _{\mathcal{H}}$ & $-0.030$ & $-0.092$ & $0.367$ & $0.306$ & $0.838$
& $0.862$ & $-0.086$ & $0.383$ & $0.318$ & $0.849$ & $0.864$ \\ 
$b_{\mathcal{H}}$ & $-0.048$ & $-0.115$ & $0.529$ & $0.292$ & $0.847$ & $%
0.872$ & $-0.108$ & $0.569$ & $0.299$ & $0.850$ & $0.870$ \\ 
$\alpha $ & $0.122$ & $0.096$ & $0.070$ & $0.318$ & $0.824$ & $0.859$ & $%
0.098$ & $0.070$ & $0.322$ & $0.832$ & $0.855$ \\ \hline
$\sigma _{\mathcal{L}}=0.1$ &  &  &  &  &  &  &  &  &  &  &  \\ \hline
$b_{\mathcal{L}}$ & $-0.902$ & $-0.794$ & $0.541$ & $0.214$ & $0.850$ & $%
0.858$ & $-0.736$ & $0.501$ & $0.223$ & $0.840$ & $0.854$ \\ 
$\delta _{\mathcal{H}}$ & $-0.078$ & $-0.120$ & $0.342$ & $0.296$ & $0.886$
& $0.897$ & $-0.113$ & $0.362$ & $0.311$ & $0.874$ & $0.887$ \\ 
$b_{\mathcal{H}}$ & $-0.092$ & $-0.140$ & $0.504$ & $0.311$ & $0.871$ & $%
0.887$ & $-0.131$ & $0.547$ & $0.296$ & $0.864$ & $0.886$ \\ 
$\alpha $ & $0.121$ & $0.093$ & $0.068$ & $0.331$ & $0.832$ & $0.873$ & $%
0.095$ & $0.069$ & $0.345$ & $0.831$ & $0.872$ \\ \hline
$\sigma _{\mathcal{L}}=0.02$ &  &  &  &  &  &  &  &  &  &  &  \\ \hline
$b_{\mathcal{L}}$ & $-0.951$ & $-0.814$ & $0.540$ & $0.217$ & $0.845$ & $%
0.847$ & $-0.749$ & $0.498$ & $0.211$ & $0.829$ & $0.842$ \\ 
$\delta _{\mathcal{H}}$ & $-0.092$ & $-0.113$ & $0.336$ & $0.333$ & $0.876$
& $0.891$ & $-0.107$ & $0.356$ & $0.324$ & $0.876$ & $0.890$ \\ 
$b_{\mathcal{H}}$ & $-0.110$ & $-0.138$ & $0.497$ & $0.314$ & $0.868$ & $%
0.873$ & $-0.131$ & $0.542$ & $0.298$ & $0.864$ & $0.879$ \\ 
$\alpha $ & $0.120$ & $0.092$ & $0.068$ & $0.338$ & $0.828$ & $0.878$ & $%
0.094$ & $0.068$ & $0.334$ & $0.837$ & $0.877$ \\ \hline
\end{tabular}

\end{center}

\footnotesize{ The table reports
the true value of the parameters (\textquotedblleft true\textquotedblright), the corresponding average estimates (\textquotedblleft estim\textquotedblright) over
1,000 Monte Carlo simulations, their standard deviations (\textquotedblleft sd\textquotedblright), the coverage rates
of 90\% confidence intervals based on 2 asymptotic approximations
(\textquotedblleft asy1\textquotedblright\ for one-way and \textquotedblleft
asy2\textquotedblright\ for two-way cluster-robust standard errors), and the
coverage rates of 90\% bootstrap confidence intervals (\textquotedblleft boot\textquotedblright). The results are
reported for 3 values of $\sigma _{\mathcal{L}}$ (the standard deviation of
the low-frequency component $\mathcal{L}$) and 2 values of $q$ for the MWq
method, which is used to estimate the latent $\mathcal{L}$.}
\end{table}

\newpage
\clearpage

\baselineskip=11.0pt
\bibliography{climate}

\end{document}